\documentclass[aps,twocolumn,prl,superscriptaddress,amsmath,showpacs,tightenlines]{revtex4-1}
\usepackage{amssymb}
\usepackage{braket}
\usepackage{amsmath}
\usepackage{graphicx}
\usepackage{subfigure}
\usepackage{natbib}
\usepackage{epsfig}
\usepackage{amsfonts}
\usepackage{mathrsfs}
\usepackage{CJK}
\usepackage{xcolor}
\usepackage{bm}
\usepackage[colorlinks,linkcolor=blue,citecolor=blue]{hyperref}%
\usepackage[english]{babel}
\usepackage{microtype}
\usepackage[toc,page,title,titletoc,header]{appendix}
\begin{document}
\title{Weakly Driven and Finite Detuning Boundary Time Crystals Enabled by Low-Dissipation Dynamical Channels}
\author{Xiang Guo}
\affiliation{Center for Quantum Sciences and School of Physics, Northeast Normal University,
Changchun 130024, China}
\author{Xiaojun Zhang}
\affiliation{Center for Quantum Sciences and School of Physics, Northeast Normal University,
Changchun 130024, China}
\author{Zhihai Wang}
\email{wangzh761@nenu.edu.cn}
\affiliation{Center for Quantum Sciences and School of Physics, Northeast Normal University, Changchun 130024, China}
\begin{abstract}
Spontaneous breaking of continuous time-translation symmetry in driven-dissipative systems gives rise to boundary time crystals (BTCs), characterized by persistent oscillations sustained by coherent driving and collective dissipation. Conventional BTCs, however, typically require strong driving and exact atom-drive resonance, imposing stringent constraints on their realization. Here we consider two atomic ensembles coupled to a common Markovian reservoir and show that shared dissipation organizes dissipation-free and low-dissipation modes into dynamically accessible low-dissipation channels, enabling BTCs under weak driving and finite detuning. Finite detuning further selects a unique stable limit cycle from an initial-state-dependent family of oscillatory trajectories. Our results establish low-dissipation dynamical channels as a route to robust BTCs under relaxed driving and resonance conditions.
\end{abstract}
\maketitle

\emph{Introduction.}---Spontaneous symmetry breaking is a cornerstone of modern physics, underlying the emergence of a broad class of ordered phases~\cite{FS2005}.
Time crystals (TCs) provide a remarkable nonequilibrium extension of this concept, where time-translation symmetry is spontaneously broken and persistent temporal oscillations emerge.
Although Wilczek's original proposal~\cite{FW2012} was later ruled out for equilibrium systems by no-go theorems~\cite{PB2013,HW2015}, subsequent studies revealed that TCs can arise in nonequilibrium settings.
In periodically driven systems, the breaking of discrete time-translation symmetry leads to discrete time crystals (DTCs)~\cite{DV2016,VK2016,SC2017,JZ2017,NY2017,DV2017,WW2017,ZG2018,SP2018,JR2018,DV2020,NY2020,AK2021,JR2021,XM2022}, whereas driven-dissipative systems provide a platform for continuous time crystals (CTCs)~\cite{KS2018,FI2018x,SA2018,BB2019,AR2020,LF2021,PK2022,TL2023,VM2023,XW2024,FR2025_2,ZW2025_2,JW2026}, characterized by spontaneous breaking of continuous time-translation symmetry.
Recent experimental demonstrations in trapped ions, NV-center spin ensembles, superfluid $^3$He, atom-cavity systems, photonic metamaterials, and Rydberg gases have further established TCs as experimentally accessible nonequilibrium phenomena~\cite{JZ2017,SC2017,SP2018,JR2018,AK2021,JR2021,XM2022,SA2018,PK2022,TL2023,XW2024}.

Boundary time crystals (BTCs), first proposed by F. Iemini \emph{et al.}~\cite{FI2018x}, constitute a paradigmatic class of dissipative CTCs in open quantum systems. In the original BTC model, a boundary ensemble of identical spin-$1/2$ particles is coherently driven while collectively coupled to a Markovian bulk reservoir, where the competition between coherent driving and collective dissipation gives rise to persistent oscillations of the boundary order parameter.

Despite its conceptual importance, the original BTC requires strong resonant driving to overcome collective dissipation. Strong driving can induce heating, decoherence, and technical noise~\cite{MR2017,AR2020_2,NQ2016,TA1997,ML2022}, while laser-frequency drift~\cite{JM2019,SJ2025} makes exact resonance difficult to maintain. Moreover, the oscillation amplitude is strongly initial-state dependent. These limitations motivate the search for BTCs that persist under weak driving and finite detuning while remaining robust against initial-state variations.

In this Letter, we demonstrate that these limitations can be substantially relaxed in a minimal system of two atomic ensembles collectively coupled to a common Markovian reservoir, with only one ensemble coherently driven. Interference between emissions from the two ensembles generates a family of dissipation-free and low-dissipation modes. At zero detuning, the resulting BTC exists for arbitrarily weak nonzero driving, while at finite detuning it persists over a broad weak-driving regime. Remarkably, finite detuning converts an initial-state-dependent family of oscillatory trajectories into a unique stable limit cycle, thereby enhancing the robustness of the time-crystalline dynamics.

To uncover the underlying mechanism, we map both the present and conventional BTC dynamics~\cite{FI2018x,LF2021,VM2023} onto wave-packet motion on effective lattices. In the two-ensemble system, dissipation-free and low-dissipation modes are dynamically organized into accessible closed channels confined to the low-dissipation region, whereas the conventional single-ensemble BTC necessarily explores strongly dissipative regions. This comparison shows that low-dissipation modes alone are insufficient for weak-driving BTCs; rather, they must form dynamically accessible channels that sustain the balance between coherent motion and dissipation. This mechanism differs from the synchronization-based non-resonant BTC of Ref.~\cite{JW2026}, which is associated
with self-sustained oscillators in the undriven limit, and the time-crystalline character is further confirmed by persistent oscillation of two-time correlations
in the thermodynamic limit~\cite{HW2015}.

\emph{Model.}---We consider a system consisting of two atomic ensembles, each containing $N_a$ identical spin-$1/2$ particles with transition frequency $\omega_a$. The atomic ensembles are collectively coupled to a common Markovian reservoir, while only the first ensemble is coherently driven by an external field with frequency $\omega_p$ and driving strength $\eta$. Under the Born--Markov approximation, the dynamics of the reduced density matrix is governed by the Lindblad master equation
\begin{align}
\frac{d\rho}{dt}=\mathcal{L}\rho
=&-i[\Delta(S_{z}^{(1)}+\alpha S_{z}^{(2)})+\eta S_{x}^{(1)},\rho]\nonumber\\
&+\frac{\gamma}{2S}
\left(
2L\rho L^{\dagger}
-L^{\dagger}L\rho
-\rho L^{\dagger}L
\right),
\label{ME1}
\end{align}
where $\Delta=\omega_a-\omega_p$ denotes the detuning between the atomic transition and the driving field. The collective spin operators are defined as $S_{\mu}^{(i)}=\sum_{j=1}^{N_a}\sigma_{\mu,j}^{(i)}/2$ $(\mu=x,y,z)$, where $\sigma_{\mu,j}^{(i)}$ denotes the Pauli operator acting on the $j$th spin of the $i$th ensemble. The dissipation is described by the jump operator $L=S_-^{(1)}+\alpha e^{-i\phi}S_-^{(2)}$, where $0\leq\phi<2\pi$ and $S_{\pm}^{(i)}=\sum_{j=1}^{N_a}\sigma_{\pm,j}^{(i)}$
are the collective spin raising and lowering operators of the $i$th ensemble.
For the present model, the phase $\phi$ can be absorbed into a local rotation of the second ensemble and therefore does not affect the dynamics discussed below. We thus set $\phi=\pi$ without loss of generality in the main text.
 For $\alpha=0$, the model reduces to a single atomic ensemble and recovers the BTC scenario proposed in Ref.~\cite{FI2018x}. For $\alpha=1$, the common reservoir induces both intra-ensemble superradiant decay and inter-ensemble dissipative coupling.

\begin{figure}
 \includegraphics[width=1\columnwidth]{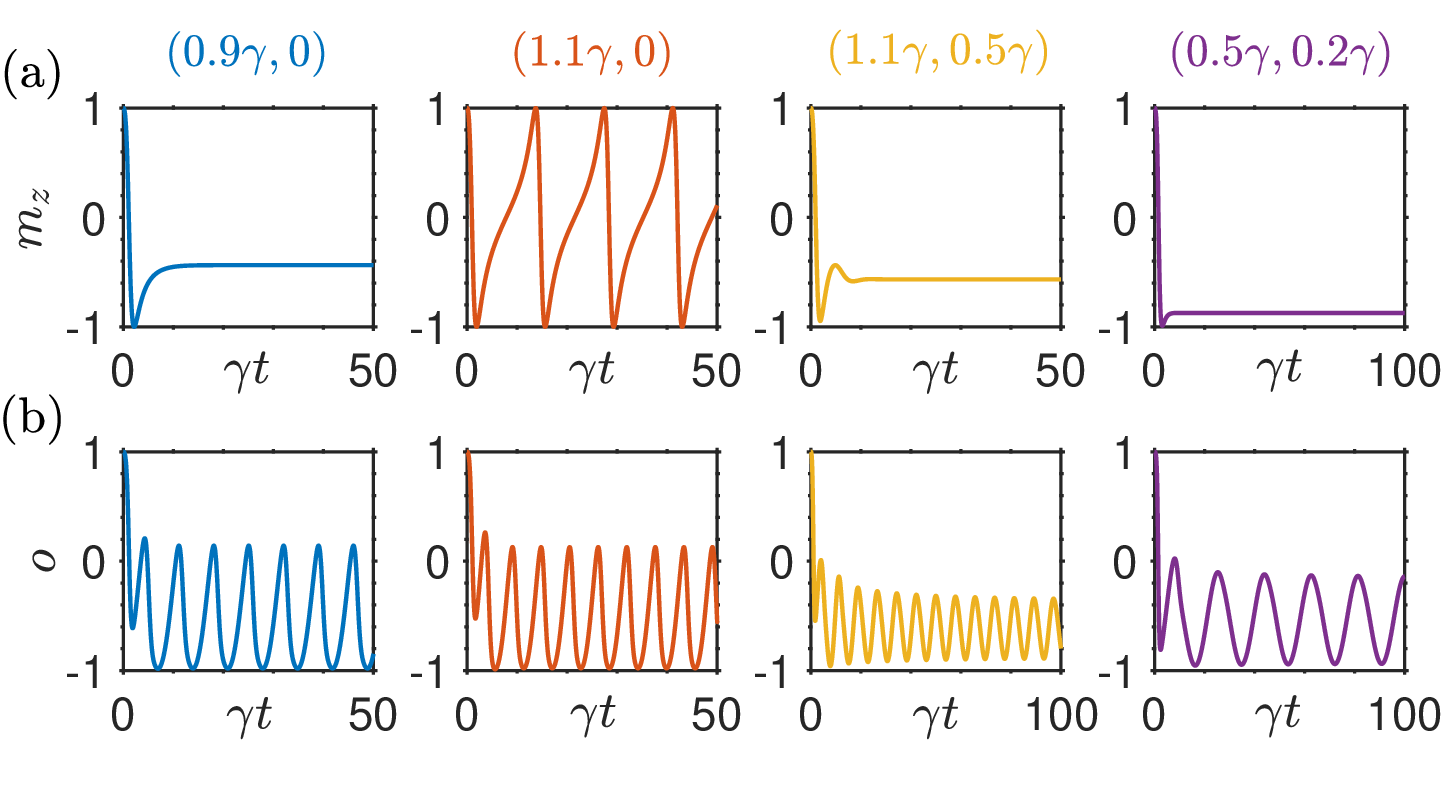}
  \caption{(a) and (b) The mean-field evolution of the order parameter under different $(\eta,\Delta)$ for $\alpha=0$ and $\alpha=1$, respectively. The initial state is set as $m_{x,y}=0,m_{z}=1$ in panel (a) and $m_{x,y}^{(1,2)}=0,m_{z}^{(1,2)}=1$ in panel (b).}
  \label{contrast_BTC}
\end{figure}

\emph{Mean-field analysis.}---We first analyze the system within the mean-field approximation, which is valid  in the thermodynamic limit $N_a\rightarrow\infty$. The mean-field equations are derived from Eq.~(\ref{ME1}) by factorizing operator correlations as $\langle AB\rangle\approx\langle A\rangle\langle B\rangle$. Their explicit form is provided in the Supplementary Material (SM)~\cite{SM}.

We begin with the single-ensemble case ($\alpha=0$), which reduces to the model studied in Ref.~\cite{FI2018x}. As shown in Fig.~\ref{contrast_BTC}(a), the order parameter
$m_z=\langle S_z^{(1)}\rangle/S$ exhibits persistent oscillations only for strong resonant driving ($\eta>\gamma$ and $\Delta=0$). The oscillations are absent either in the weak-driving regime ($\eta<\gamma$)
or at finite detuning ($\Delta\neq0$), reflecting the restrictive conditions for BTC formation in this setting.

The situation changes qualitatively when a second undriven ensemble is coupled to the same reservoir ($\alpha=1$). As shown in Fig.~\ref{contrast_BTC}(b), the order parameter
$o=\sum_{i=1,2}\langle S_z^{(i)}\rangle/(2S)$ displays periodic oscillations in both the weak- and strong-driving regimes. More remarkably, the BTC persists under finite detuning, demonstrating that neither strong driving nor exact resonance is required for its emergence. Figure~S1 (SM~\cite{SM}) further reveals the stabilizing role of detuning. At $\Delta=0$, the BTC oscillation amplitude depends on the initial state, whereas for $\Delta\neq0$, different initial states converge to the same limit cycle, differing only by an overall phase shift.

\emph{Effective lattice picture.}---To uncover the origin of the BTC under weak driving and finite detuning in two-ensemble scheme, we analyze the dissipative structure of the system through an effective lattice picture. We show that dissipation-free and low-dissipation modes provide protected dynamical channels, within which a stable balance between coherent interactions and dissipation sustains persistent oscillations.

Since each atomic ensemble remains in the fully symmetric subspace, its Hilbert space is spanned by the Dicke manifold $\mathcal{H}_{S}^{(i)}$ ($i=1,2$) with total spin $S=N_a/2$. The Hilbert space of the composite system is therefore $\mathcal{H}_{S}^{(1)}\otimes\mathcal{H}_{S}^{(2)}=\bigoplus_{j=0}^{2S}\mathcal{T}_{j}$, according to the Clebsch-Gordan (CG) decomposition~\cite{AC1866x,HW1931x,EP1959x}. Here, each subspace $\mathcal{T}_{j}$ is spanned by the angular-momentum basis $\{\ket{j,m}\}$, which are the common eigenstates of ${\bm J}^{2}$ and $J_z$, with ${\bm J}=\bm S^{(1)}+\bm S^{(2)}$ denoting the total angular momentum.

\begin{figure}
  \includegraphics[width=0.8\columnwidth]{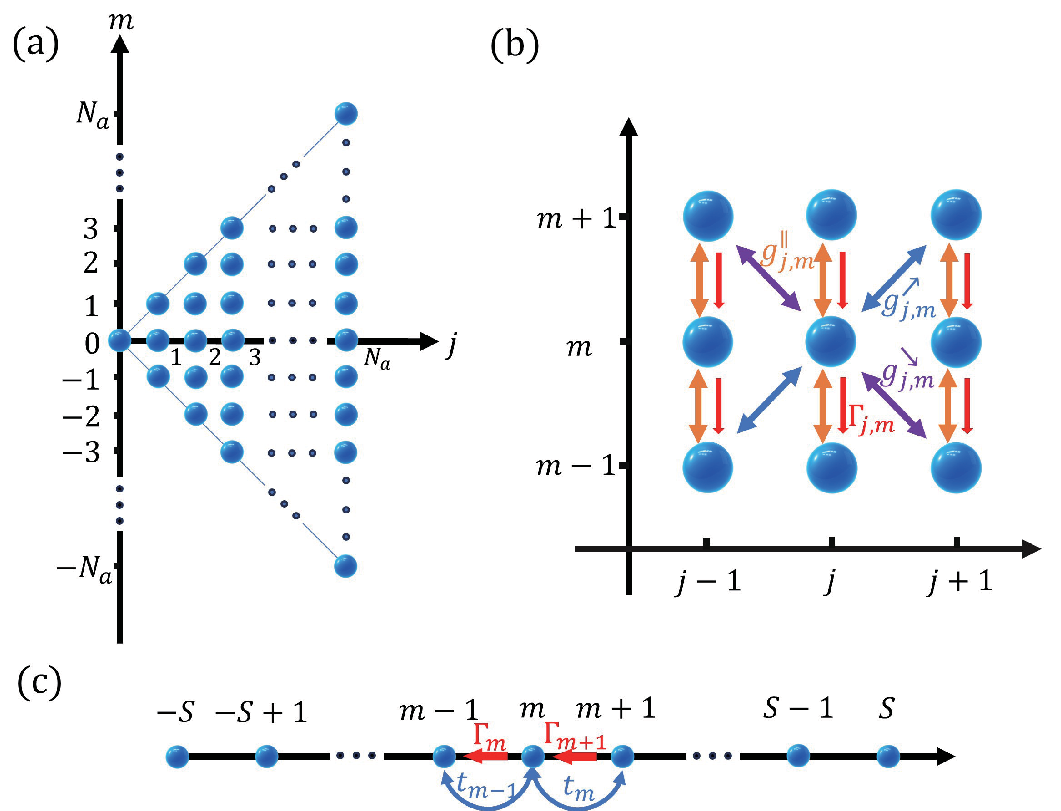}\\
  \caption{Effective lattice picture of the system.
(a) Triangular lattice for $\alpha=1$, where each lattice site represents a basis state $\ket{\varphi_{j,m}}$.
(b) Local interaction structure around an arbitrary bulk site $\ket{\varphi_{j,m}}$ of the triangular lattice. The double-headed arrows denote the three types of coherent intersite couplings, characterized by $g_{j,m}^{\parallel}$, $g_{j,m}^{\nearrow}$, and $g_{j,m}^{\searrow}$, while the red arrows indicate the dissipative transitions with rate $\Gamma_{j,m}$.
(c) One-dimensional lattice picture for $\alpha=0$, with coherent nearest-neighbor couplings $t_m$ and cascaded dissipative transitions at rates $\Gamma_m$.}
  \label{Model_MT}
\end{figure}

To eliminate the $\phi$ dependence of the dissipation operator, we introduce the unitary transformation $U=e^{i\phi S_z^{(2)}}$, such that the states $\{\ket{\varphi_{j,m}}:=U\ket{j,m}\}$ form a complete basis of the composite system. As illustrated in Fig.~\ref{Model_MT}(a), these basis states can be naturally arranged into an effective triangular lattice. In this representation, the many-body evolution is mapped onto the motion of a wave packet over the lattice, with the dissipative and coherent inter-site processes fully determined by the Lindblad master equation Eq.~(\ref{ME1}).

First, we examine how the dissipative process redistributes the populations across the lattice. The population of each lattice site,
$P_{j,m}=\langle \varphi_{j,m}|\rho(t)|\varphi_{j,m}\rangle$,
obeys
\begin{equation}
\dot{P}_{j,m}=-\Gamma_{j,m} P_{j,m}+\Gamma_{j,m+1}P_{j,m+1}
\label{Pjm}
\end{equation}
where the site-dependent dissipative transition rate is
$\Gamma_{j,m}=\gamma(j+m)(j-m+1)/S$,
with the explicit derivation being presented in the SM~\cite{SM}. Here, the first term in Eq.~(\ref{Pjm}) describes the loss from $\ket{\varphi_{j,m}}$ through the transition
$\ket{\varphi_{j,m}}\rightarrow \ket{\varphi_{j,m-1}}$,
whereas the second term describes the population gain through
$\ket{\varphi_{j,m+1}}\rightarrow \ket{\varphi_{j,m}}$.
Consequently, the Lindblad dissipation generates a cascade of unidirectional transitions along each column of the triangular lattice, as indicated by the red arrows in Fig.~\ref{Model_MT}(b).

Importantly, the transition rate sensitively depends on the lattice sites through the quantum number $m$. The dissipation rate $\Gamma_{j,m}$ is strongly suppressed when $m\approx-j$ and vanishes exactly at the lower boundary ($m=-j$). Consequently, the lower boundary of the triangular lattice hosts dissipation-free modes, while its vicinity supports low-dissipation modes. As will be shown below, these modes provide the protected dynamical channels through which a stable balance between coherent interactions and dissipation can be established even under weak driving and finite detuning.

Next, we turn to the coherent inter-site couplings. In the basis
$\{\ket{\varphi_{j,m}}\}$, the Hamiltonian
$H=\Delta J_z+\eta S_x^{(1)}$ can be decomposed as
$H=H_f+H_{\parallel}+H_{\nearrow}+H_{\searrow}$, where
\begin{align}
H_f
&=\sum_{j=0}^{N_a}\sum_{m=-j}^{j}
\Delta m\,X_{(j,m),(j,m)},\nonumber\\
H_{\parallel}
&=\eta\sum_{j=0}^{N_a}\sum_{m=-j}^{j-1}
g_{j,m}^{\parallel}
X_{(j,m+1),(j,m)}
+{\rm H.c.},\nonumber\\
H_{\nearrow}
&=\eta\sum_{j=0}^{N_a-1}\sum_{m=-j}^{j}
g_{j,m}^{\nearrow}
X_{(j+1,m+1),(j,m)}
+{\rm H.c.},\nonumber\\
H_{\searrow}
&=\eta\sum_{j=0}^{N_a-1}\sum_{m=-j}^{j}
g_{j,m}^{\searrow}
X_{(j+1,m-1),(j,m)}
+{\rm H.c.}.
\end{align}
Here,
$X_{(j,m),(j',m')}
\equiv
\ket{\varphi_{j,m}}\bra{\varphi_{j',m'}}$
denotes the transition operator between two effective lattice sites.
Here, $H_f$ describes the onsite energy of each effective lattice site. Furthermore, as illustrated by the double-headed arrows in Fig.~\ref{Model_MT}(b), $H_{\parallel}$ couples nearest-neighbor sites within the same column, while $H_{\nearrow}$ and $H_{\searrow}$ connect next-nearest-neighbor sites in adjacent columns through upward and downward diagonal hoppings, respectively. The explicit expressions of the coupling coefficients $g_{j,m}^{\beta},\,(\beta=\parallel,\nearrow,\searrow)$ are presented in the SM~\cite{SM}.

For an arbitrary quantum state $\ket{\psi(t)}=\sum_{j=0}^{N_a}
\sum_{m=-j}^{j}\psi_{j,m}(t)\ket{\varphi_{j,m}}$, the probability amplitudes $\psi_{j,m}(t)$ can be viewed as a wave packet evolving on the effective triangular lattice.
Its dynamics are governed by the position-dependent coherent hoppings, dissipation and the initial condition.

In the thermodynamic limit ($N_a\rightarrow\infty$) demanded for BTC formation, the triangular lattice becomes continuous, with the lattice coordinates parameterized by
$x=j/N_a$ and $z=m/N_a$.
Using the semiclassical and narrow-wave-packet approximations, we derive the canonical equations governing the wave-packet center in phase space (see Eq.~(S28) of the SM~\cite{SM}). Remarkably, the order parameter is directly determined by the wave-packet center $(x(t),z(t))$, with $o(t)=z(t)$, establishing a direct correspondence between the semiclassical dynamics and the BTC oscillations.

To characterize the wave-packet motion, we introduce two orthogonal coordinates, $v=(x+z)/\sqrt{2},\,u=(x-z)/\sqrt{2}$.
Since the lower and upper boundaries of the triangular lattice satisfy $z=-x$ and $z=x$ respectively, $v$ ($u$) measures the perpendicular distance from the wave-packet center to the lower (upper) boundary.

\begin{figure}
 \includegraphics[width=0.5\columnwidth]{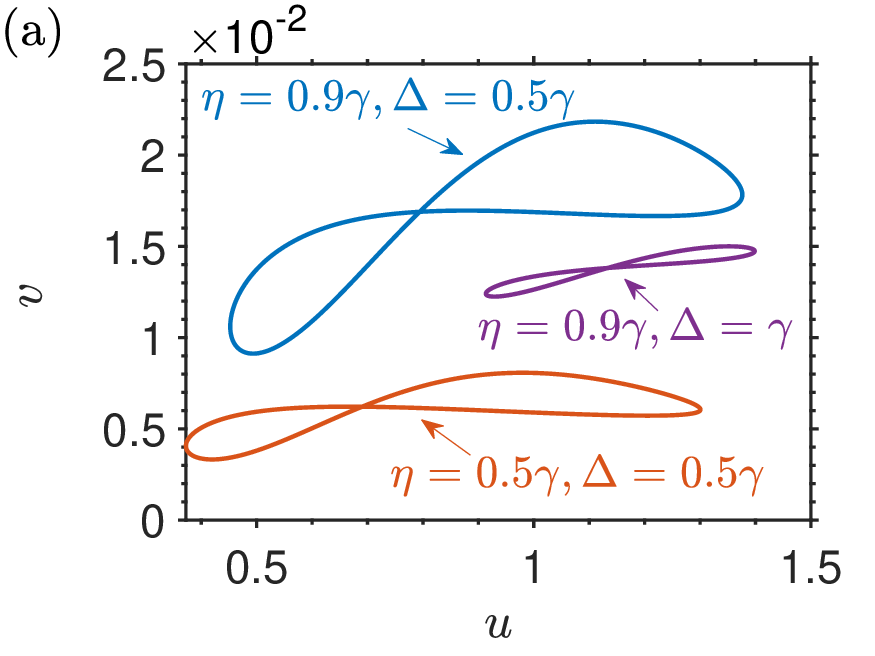}\includegraphics[width=0.5\columnwidth]{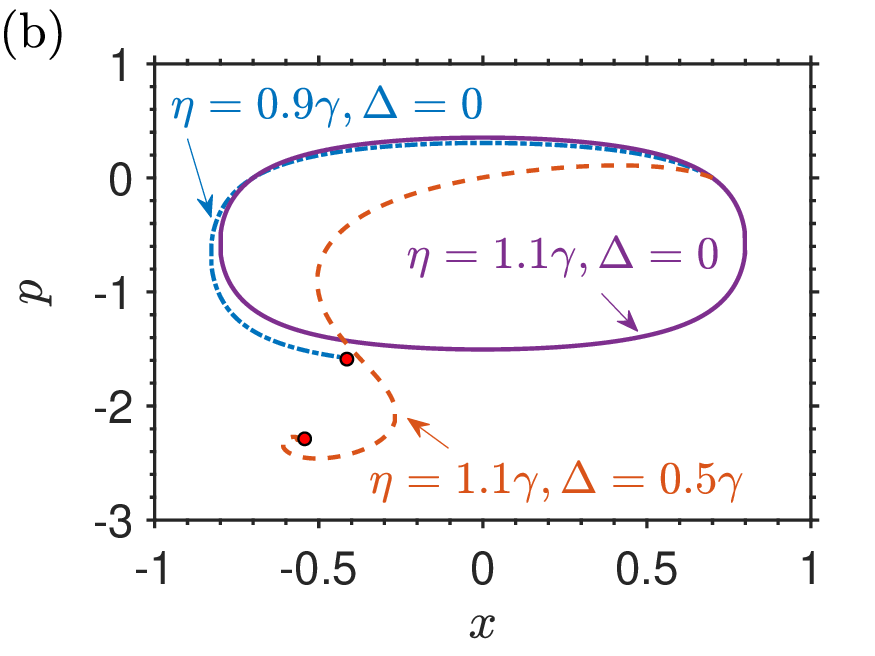}
  \caption{(a) The semiclassical phase space limit cycles projected onto the (a) $u$-$v$ and (b) $x$-$p$ plane for $\alpha=1$ and $\alpha=0$, respectively.  The red circles in (b) mark the fixed points approached after long-time evolution under parameters outside the BTC phase. All curves in panel (b) adopt the identical initial condition $x(0)=0.7,\ p(0)=0$.}
  \label{traj_MT}
\end{figure}

For $\eta<\gamma$ and $\Delta\neq0$, trajectories starting from different initial conditions converge to the same closed orbit, as shown in Fig.~\ref{traj_MT}(a), consistent with Fig.~S1 of the SM~\cite{SM}. Remarkably, this orbit remains confined near the low-dissipation boundary $v=0$. In the effective-lattice picture, hopping away from this boundary mediated by $H_{\parallel}$ and $H_{\nearrow}$ is suppressed by the enhanced dissipation and detuning-induced energy mismatch, whereas $H_{\searrow}$ sustains excitation exchange along the low-dissipation boundary. The resulting motion exhibits pronounced oscillations along the $u$ direction, which are directly transferred to the order parameter through $o(t)=z(t)=[v(t)-u(t)]/\sqrt{2}$. Thus, dynamically accessible low-dissipation channels can sustain BTCs without strong driving and exact resonance.

The evolution of the trajectories with $\eta$ and $\Delta$ in Fig.~\ref{traj_MT}(a) further supports this picture. Reducing $\eta$ weakens the coherent hopping relative to dissipation, shifting the dynamical balance toward smaller $v$, where dissipation is further suppressed. Increasing $\Delta$ similarly suppresses effective excitation exchange by enlarging the energy mismatch between coherently coupled lattice sites, driving the trajectory closer to the lower boundary and reducing its oscillation amplitudes along both $u$ and $v$.

For comparison, we revisit the conventional single-ensemble case ($\alpha=0$)~\cite{FI2018x}. Here the effective triangular lattice reduces to the one-dimensional chain shown in Fig.~\ref{Model_MT}(c), with each site corresponding to a Dicke state $\ket{S,m}$. Neighboring sites are coherently coupled by the position-dependent hopping
$t_m=\sqrt{(S-m)(S+m+1)}$, while collective dissipation induces cascaded transitions at the rate
$\Gamma_m=\gamma(S+m)(S-m+1)/S$. The dissipation therefore vanishes at the endpoint $m=-S$ and remains weak in its vicinity.

In the thermodynamic limit, we introduce the continuous coordinate $x=m/S$ and, within the narrow-wave-packet approximation, identify the wave-packet center with the order parameter, $x(t)=m_z(t)$; its dynamics are governed by Eq.~(S49) of the SM~\cite{SM}. Since the dissipation-free endpoint corresponds to $x=-1$, we define $v(t)=1+x(t)$ as the distance from this endpoint. The semiclassical BTC trajectory, shown by the purple curve in Fig.~\ref{traj_MT}(b), extends far from $v=0\,(x=-1)$ and necessarily explores regions of substantial dissipation. This demonstrates that the mere presence of dissipation-free and low-dissipation modes is insufficient for BTC formation; rather, persistent oscillations require a dynamically accessible channel in which coherent hopping can balance dissipation. For $\eta<\gamma$ and $\Delta=0$, the hopping is too weak to maintain such a balance, and the trajectory relaxes to the fixed point marked by the red endpoint of the blue dash-dotted curve, whose $x$ coordinate agrees with the steady-state value of $m_z$ in Fig.~\ref{contrast_BTC}(a). In sharp contrast, the two-dimensional lattice supports an accessible closed channel confined near the low-dissipation boundary, thereby allowing BTCs to persist under weak driving.

This picture also clarifies why BTCs in the single-ensemble system are restricted to exact resonance. Because the BTC trajectory necessarily traverses high-dissipation regions, persistent oscillations require sufficiently strong coherent hopping to compensate for the associated dissipation. At finite detuning, the resulting intersite energy mismatch suppresses coherent hopping and destroys the dynamical balance required to sustain the closed trajectory. The trajectory therefore loses its closed-orbit character and relaxes to a stable fixed point, marked by the red endpoint of the orange dashed curve in Fig.~\ref{traj_MT}(b). The corresponding $x$ coordinate agrees with the steady-state value of $m_z$ obtained from the mean-field dynamics in Fig.~\ref{contrast_BTC}(a).

\begin{figure}
 \includegraphics[width=0.5\columnwidth]{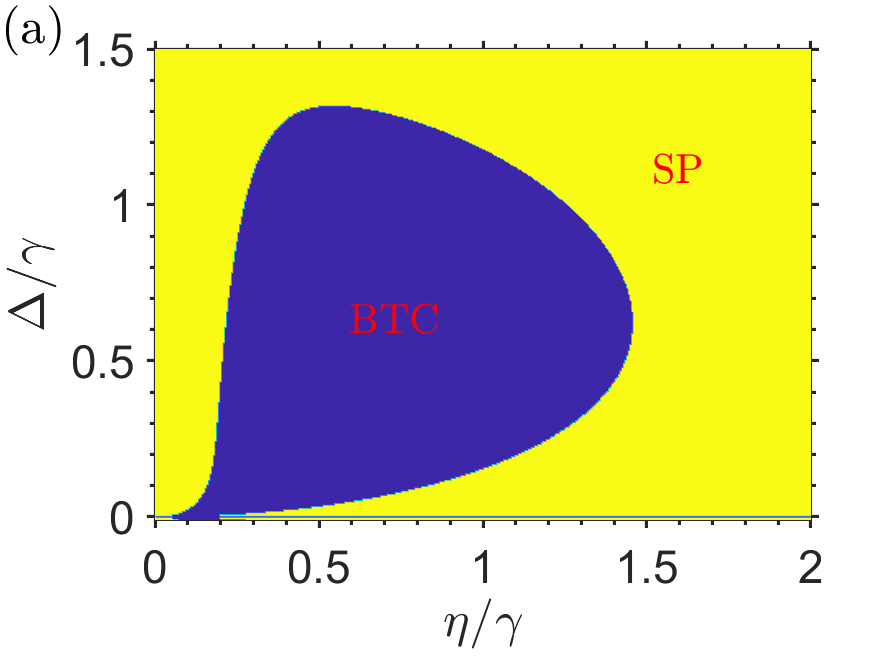}\includegraphics[width=0.5\columnwidth]{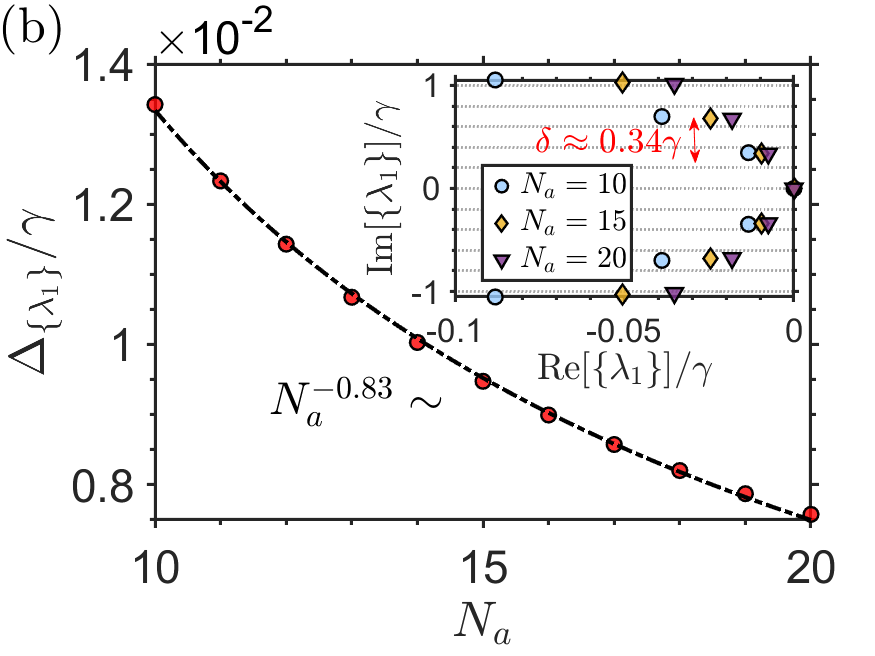}\\
 \includegraphics[width=0.5\columnwidth]{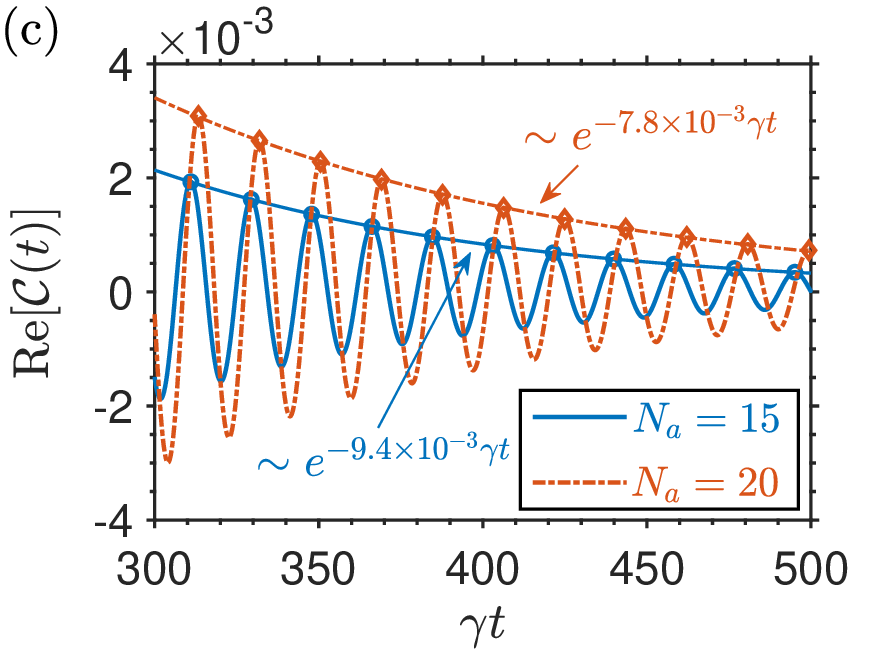}\includegraphics[width=0.5\columnwidth]{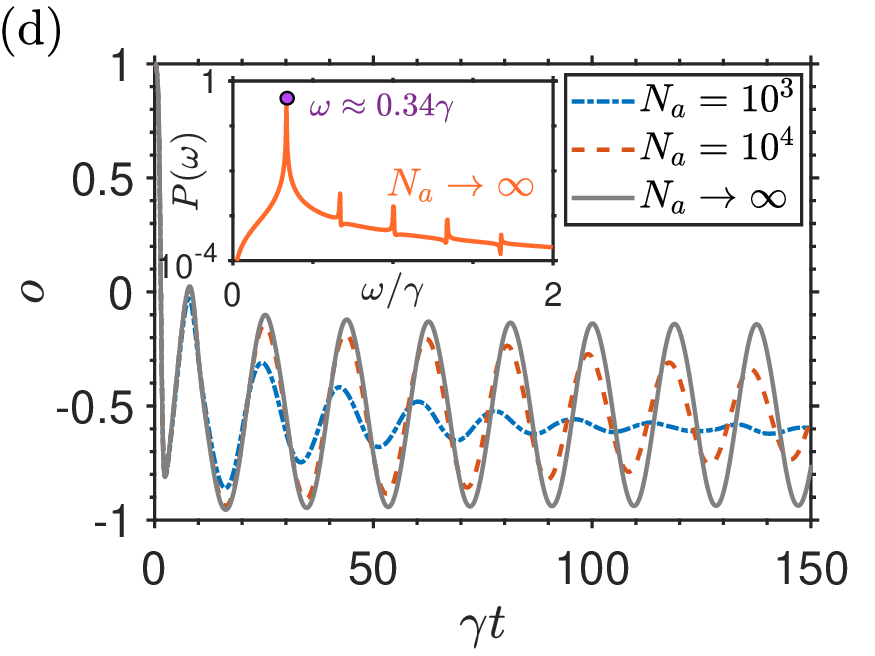}
  \caption{(a) Mean-field phase diagram in the $\eta-\Delta$ plane. (b) Energy gap of the first eigenvalue branch as a function of $N_a$ with power-law fitting. Inset of (b): First eigenvalue branches in the low-lying Liouvillian spectrum for $N_a=10, 15, 20$. (c) Long-time dynamics of real part of the two-time correlation function $\mathcal{C}(t)$ for $N_a=15$ and $20$. (d) Dynamics of the order parameter $o$ against system size. Inset of (d): Fourier spectrum of $o$ in the thermodynamic limit. In (d), $5000$ stochastic trajectories are adopted in the TWA simulation for $N_{a}=10^3$ and $N_{a}=10^4$ and the initial state is set as $m_{x,y}^{(1,2)}=0,m_{z}^{(1,2)}=1$. For panels (b), (c) and (d), we set $\Delta=0.2\gamma$ and $\eta=0.5\gamma$. }
  \label{diagram&quantum}
\end{figure}

\emph{Phase diagram and quantum analysis.}---We now determine the phase diagram by performing a linear stability analysis of the mean-field steady states. Figure~\ref{diagram&quantum}(a) shows the resulting phase diagram in the $(\eta,\Delta)$ plane, with the blue and yellow regions denoting the BTC and stable steady-state phases (SP), respectively. At resonance ($\Delta=0$), the BTC exists for any nonzero driving strength, indicating the absence of a finite driving threshold. Remarkably, this weak-driving BTC is not restricted to resonance: it extends over a broad region at finite detuning, with $\eta<\gamma$. Moreover, throughout the detuned BTC regime, the system approaches a unique stable limit cycle, consistent with the dynamical picture established above.

To further uncover the quantum signature of the BTC in our scheme, we turn to the Liouvillian spectrum. As a representative example, we consider a parameter point deep inside the BTC phase with weak driving and finite detuning,
$\eta=0.5\gamma$ and $\Delta=0.2\gamma$.
The Liouvillian eigenvalues are organized into a sequence of ordered branches $\{\lambda_k\}$. As the system size increases from $N_a=10$ to $20$, the low-lying branches move progressively toward the imaginary axis (Fig.~S3 in the SM~\cite{SM}), which is a universal spectral feature of dissipative systems approaching a stable limit cycle~\cite{SD2025}.

The first eigenvalue branch $\{\lambda_1\}$, shown in the inset of Fig.~\ref{diagram&quantum}(b), exhibits a characteristic parabolic structure associated with quantum diffusion~\cite{SD2025}. More importantly, the eigenvalues are nearly equally spaced along the imaginary axis with an approximately constant interval
$\delta\simeq0.34\gamma$.
This uniform spacing directly determines the oscillation frequency of the BTC in the thermodynamic limit. The main panel of Fig.~\ref{diagram&quantum}(b) further shows that the corresponding Liouvillian gap decreases algebraically with the system size,
$\Delta_{\{\lambda_1\}}\sim N_a^{-0.83}\gamma$. Therefore, the Liouvillian gap asymptotically closes in the thermodynamic limit $N_a\rightarrow\infty$, providing the spectral foundation for persistent oscillations in the BTC phase.

The Liouvillian spectral closure merely indicates the emergence of a family of dissipation-free oscillatory Liouvillian modes, whose overlap with the selected order parameter may vanish~\cite{HA2022}. Therefore, the BTC must be further identified through the long-time behavior of the two-time correlation function of the order-parameter fluctuations~\cite{HW2015},
\begin{align}
\mathcal{C}(t)
=
\mathrm{Tr}
\!\left[
\delta\mathcal O(t)
\delta\mathcal O(0)
\rho_{\rm ss}
\right],
\end{align}
where
$\delta\mathcal O(t)=\mathcal O(t)-\langle\mathcal O\rangle_{\rm ss}$,
$\langle\mathcal O\rangle_{\rm ss}
=\mathrm{Tr}[\mathcal O\rho_{\rm ss}]$. Here, we have chosen $\mathcal O=J_z/N_a$,
and
$\rho_{\rm ss}$ denotes the unique steady state for finite $N_a$, satisfying $\mathcal{L}(\rho_{\rm ss})=0$, where $\mathcal{L}$ denotes the Liouvillian superoperator associated
with the master equation Eq.~(\ref{ME1}). Persistent oscillations of $\mathrm{Re}[\mathcal C(t)]$ signals spontaneous breaking of continuous time-translation symmetry, which constitutes the defining characteristic of a BTC~\cite{KT2018,TC2022}.

Using the quantum regression theorem~\cite{MO1997}, we calculate
$\mathrm{Re}[\mathcal C(t)]$
for
$N_a=15$
and
$20$,
as shown in Fig.~\ref{diagram&quantum}(c).
For finite $N_a$, the long-time oscillation amplitude decays exponentially, with the decay rate governed by the Liouvillian gap of the first branch shown in Fig.~\ref{diagram&quantum}(b). Since the gap scales as $\Delta_{\{\lambda_1\}}\sim N_a^{-0.83}\gamma$, the lifetime of the oscillation diverges algebraically with the system size, implying truly persistent oscillations only in the thermodynamic limit.

The corresponding dynamics of the order parameter are shown in Fig.~\ref{diagram&quantum}(d). Since exact diagonalization becomes intractable for $N_a=10^3$ and $10^4$, we employ the truncated Wigner approximation (TWA)~\cite{JH2021} based on stochastic trajectories. The oscillation amplitude rapidly converges to the thermodynamic-limit value with increasing $N_a$.
In the limit $N_a\rightarrow\infty$, persistent oscillations emerge and are captured by the mean-field approximation, as indicated by the gray curve in Fig.~\ref{diagram&quantum}(d).

Finally, we perform Fourier analysis of the persistent oscillation. As shown in the inset of Fig.~\ref{diagram&quantum}(d), the frequency spectrum exhibits nearly equally spaced peaks. The dominant oscillation frequency, $\omega\simeq0.34\gamma$, agrees remarkably well with the spacing between the imaginary parts of neighboring eigenvalues in the first Liouvillian branch shown in Fig.~\ref{diagram&quantum}(b). Additional Liouvillian spectral analysis for the resonant weak-driving BTC ($\eta=0.9\gamma,\Delta=0$) is presented in the SM~\cite{SM}.

\emph{Conclusion and remarks.}---In this Letter, we identify a mechanism for realizing BTCs under weak driving and finite detuning, where low-dissipation modes organize into dynamically accessible channels. Our system consists of two atomic ensembles collectively coupled to a common Markovian reservoir. We show that shared dissipation relaxes the conventional constraints on BTC formation by generating low-dissipation modes with suitable dynamical accessibility. An effective lattice description reveals that these modes form protected channels supporting the dynamical balance required for BTCs, providing a route to time-crystalline dynamics beyond strong-driving and exact-resonance paradigm. Furthermore, we show that finite detuning plays a constructive role in BTC formation by selecting a unique stable limit cycle from a family of oscillatory trajectories.

Our proposal can be implemented in state-of-the-art waveguide-QED platforms based on superconducting quantum circuits. In such systems, a one-dimensional transmission line serves as a common waveguide that mediates collective dissipation between distant superconducting artificial atoms or Rydberg atomic ensembles. As shown in the SM~\cite{SM}, the master equation considered in this work can be derived microscopically by treating the waveguide as a structured reservoir, providing a realistic route toward the experimental realization of the proposed BTC.

The low-dissipation-mode mechanism and the detuning-induced stabilization of time-crystalline dynamics in this Letter provide distinguish perspectives for dissipative many-body systems and the realization of experimentally accessible time crystals.

\emph{Acknowledgments}.-This work is supported by the National Natural Science Foundation of China (Grant No. 12375010) and Quantum Science and Technology-National Science and Technology Major Project (No. 2023ZD0300700).

\emph{Data availability}.-The data that support the findings of this Letter are not publicly available but are available from the authors upon reasonable request.

\begingroup
\renewcommand{\addcontentsline}[3]{}

\endgroup

\clearpage

% Start Supplementary Material on a new page
\onecolumngrid

\renewcommand{\citenumfont}[1]{S#1}
\setcounter{page}{1}

% Reset equation, figure, table, and section numbers
\setcounter{equation}{0}
\setcounter{figure}{0}
\setcounter{table}{0}
\setcounter{section}{0}
\setcounter{secnumdepth}{3}

% Add S to equation, figure, table, section numbers
\renewcommand{\theequation}{S\arabic{equation}}
\renewcommand{\thefigure}{S\arabic{figure}}
\renewcommand{\thetable}{S\arabic{table}}
\renewcommand{\thesection}{S\arabic{section}}

% ============================================================
% Supplementary title
% ============================================================

\begin{center}

{\large\bfseries
Supplementary Material for\\[5pt]
``Weakly Driven and Finite Detuning Boundary Time Crystals Enabled by
Low-Dissipation Dynamical Channels''
}

\vspace{0.5cm}

Xiang Guo, Xiaojun Zhang, and Zhihai Wang$^{*}$

\vspace{0.2cm}

\textit{
Center for Quantum Sciences and School of Physics,\\
Northeast Normal University, Changchun 130024, China
}
\end{center}

\vspace{0.5cm}

This Supplementary Material (SM) is organized into five sections.
In Sec.~\ref{A}, we derive the mean-field equations and present the order-parameter dynamics for different initial states.
In Sec.~\ref{B}, we develop a semiclassical description of the two-ensemble system
In Sec.~\ref{C}, we apply an analogous semiclassical analysis to the single-ensemble BTC.
In Sec.~\ref{D}, we present the complete low-lying Liouvillian spectrum of the two-ensemble system and provide complementary quantum evidence for the BTC in the weak-driving and zero-detuning regime through the Liouvillian spectrum and two-time correlation function.
Finally, in Sec.~\ref{E}, we derive the master equation of our model from a microscopic waveguide-QED setup with a one-dimensional linear dispersion relation.

\tableofcontents

\vspace{3em}

\section{Mean-field approximation}\label{A}

In this section, starting from the master equation Eq.~(1) of the main text, we derive the mean-field equations in the thermodynamic limit. We then determine the phase diagram in the $\eta$--$\Delta$ plane through a stability analysis of the mean-field steady states.

The system consists of two atomic ensembles, each composed of $N_a$ identical two-level atoms with transition frequency $\omega_a$, collectively coupled to a common Markovian environment. The collective operators of the $i$th ensemble are defined as $S_{\mu}^{(i)}=(1/2)\sum_{j=1}^{N_a}\sigma_{\mu,j}^{(i)}$ with $\mu=x,y,z$, where $\sigma_{\mu,j}^{(i)}$ denotes the Pauli operator acting on the $j$th atom of the $i$th ensemble. The common environment gives rise to both collective dissipation within each ensemble and correlated dissipation between the two ensembles. In addition, a coherent field with frequency $\omega_p$ and driving strength $\eta$ is applied only to the first ensemble. The system dynamics is governed by the master equation
\begin{align}
\frac{d\rho}{dt}=\mathcal{L}
\rho=-i[\Delta(S_z^{(1)}+S_z^{(2)})+\eta S_x^{(1)},\rho]+\frac{\gamma}{2S}(2L\rho L^{\dagger}-L^{\dagger}L\rho-\rho L^{\dagger}L),
\label{MES}
\end{align}
where $\Delta=\omega_a-\omega_p$ denotes the detuning between the driving field and the atomic transition. The collective jump operator is $L=S_-^{(1)}+\alpha e^{-i\phi}S_-^{(2)}$, with $0\leq\phi<2\pi$, and $\gamma$ characterizes the collective dissipation rate.

Within the mean-field approximation, correlations between distinct operators are neglected according to
$\langle AB\rangle\approx\langle A\rangle\langle B\rangle$,
thereby yielding a closed set of nonlinear equations for the collective-operator expectation values. Defining
$m_{\mu}^{(i)}=\langle S_{\mu}^{(i)}\rangle/S$
with $\mu=x,y,z$ and $i=1,2$, the mean-field equations corresponding to Eq.~(\ref{MES}) take the form
\begin{align}
\frac{d m_{x}^{(1)}}{dt}
&=-\Delta m_{y}^{(1)}
+\gamma\Big[m_{x}^{(1)}
+\alpha\cos(\phi)m_{x}^{(2)}
-\alpha\sin(\phi)m_{y}^{(2)}\Big]m_{z}^{(1)},
\nonumber\\
\frac{d m_{y}^{(1)}}{dt}
&=\Delta m_{x}^{(1)}
-\eta m_{z}^{(1)}
+\gamma\Big[m_{y}^{(1)}
+\alpha\cos(\phi)m_{y}^{(2)}
+\alpha\sin(\phi)m_{x}^{(2)}\Big]m_{z}^{(1)},
\nonumber\\
\frac{d m_{z}^{(1)}}{dt}
&=\eta m_{y}^{(1)}
-\gamma\Big[
(m_{x}^{(1)})^2+(m_{y}^{(1)})^2
+\alpha\cos(\phi)
\big(m_{x}^{(1)}m_{x}^{(2)}+m_{y}^{(1)}m_{y}^{(2)}\big)
-\alpha\sin(\phi)
\big(m_{x}^{(1)}m_{y}^{(2)}-m_{y}^{(1)}m_{x}^{(2)}\big)
\Big],
\nonumber\\
\frac{d m_{x}^{(2)}}{dt}
&=-\alpha\Delta m_{y}^{(2)}
+\gamma\Big[
\alpha^2m_{x}^{(2)}
+\alpha\cos(\phi)m_{x}^{(1)}
+\alpha\sin(\phi)m_{y}^{(1)}
\Big]m_{z}^{(2)},
\nonumber\\
\frac{d m_{y}^{(2)}}{dt}
&=\alpha\Delta m_{x}^{(2)}
+\gamma\Big[
\alpha^2m_{y}^{(2)}
+\alpha\cos(\phi)m_{y}^{(1)}
-\alpha\sin(\phi)m_{x}^{(1)}
\Big]m_{z}^{(2)},
\nonumber\\
\frac{d m_{z}^{(2)}}{dt}
&=-\gamma\Big[
\alpha^2\big((m_{x}^{(2)})^2+(m_{y}^{(2)})^2\big)
+\alpha\cos(\phi)
\big(m_{x}^{(1)}m_{x}^{(2)}+m_{y}^{(1)}m_{y}^{(2)}\big)
-\alpha\sin(\phi)
\big(m_{x}^{(1)}m_{y}^{(2)}-m_{y}^{(1)}m_{x}^{(2)}\big)
\Big].
\label{MF}
\end{align}
Here, $\alpha=0$ and $\alpha=1$ correspond to the single- and two-ensemble configurations, respectively. A standard linear stability analysis of Eq.~(\ref{MF}) for $\alpha=1$ yields the phase diagram shown in Fig.~4(a) of the main text.

\begin{figure}
  \includegraphics[width=0.23\columnwidth]{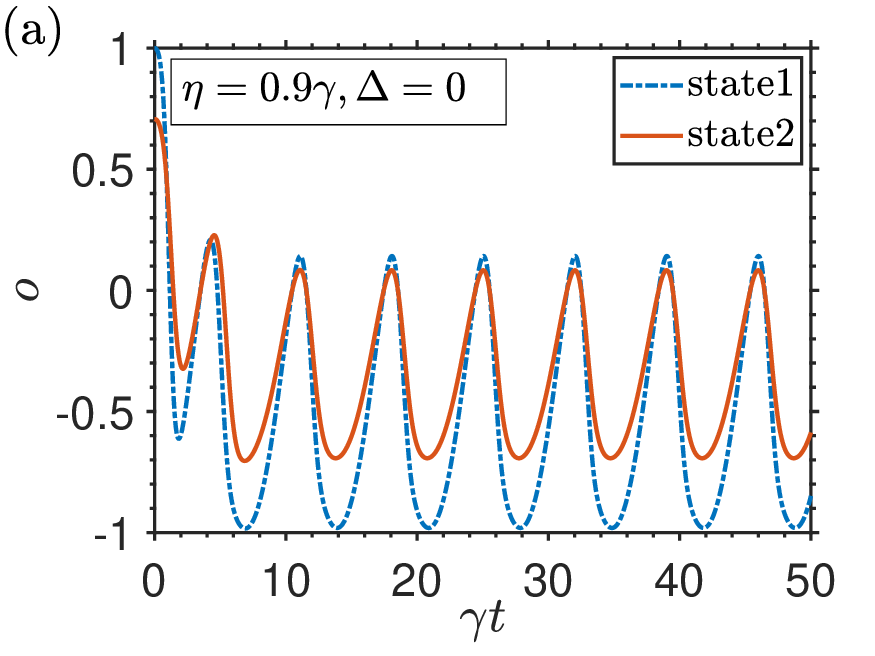}
  \includegraphics[width=0.23\columnwidth]{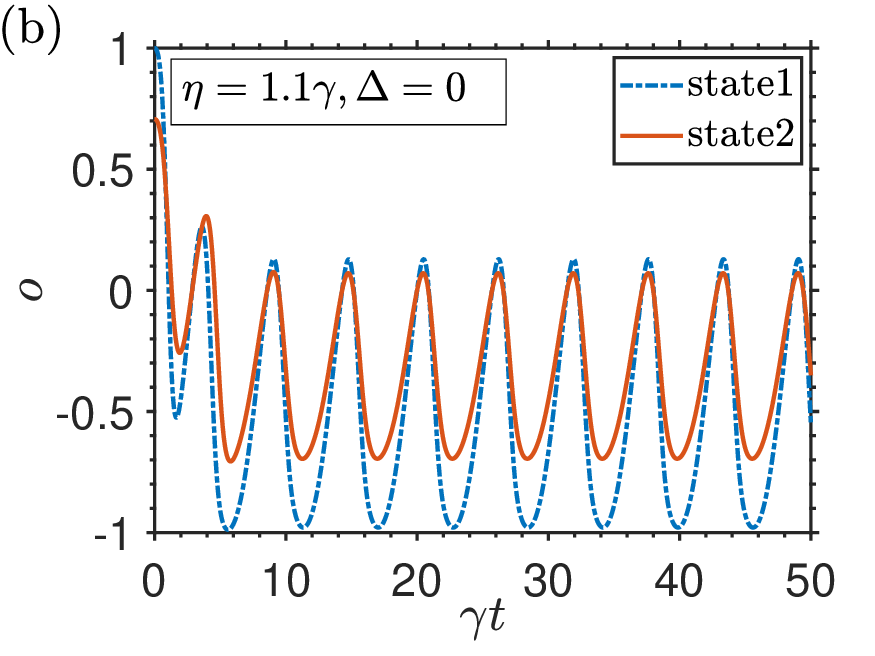}
  \includegraphics[width=0.23\columnwidth]{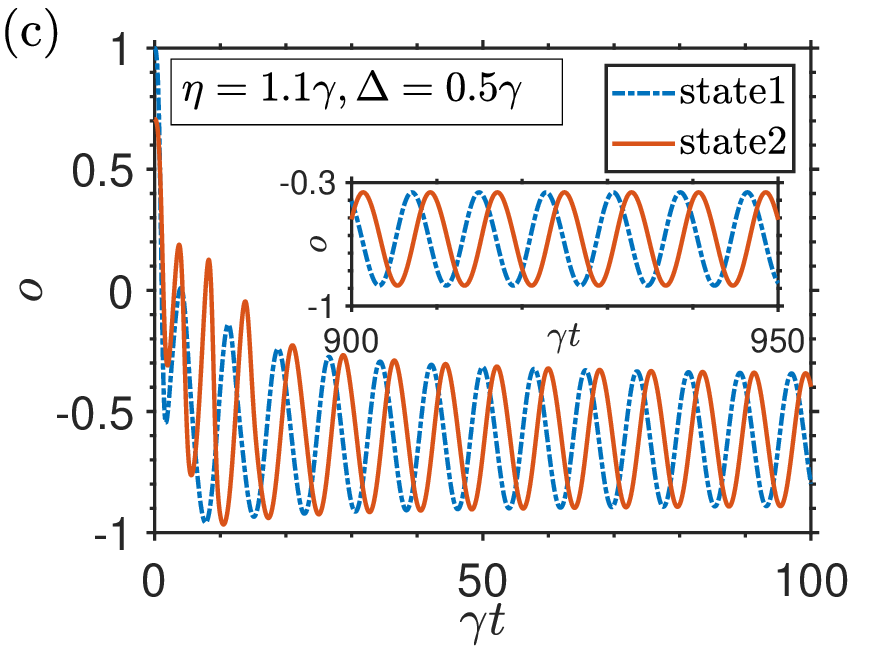}
  \includegraphics[width=0.23\columnwidth]{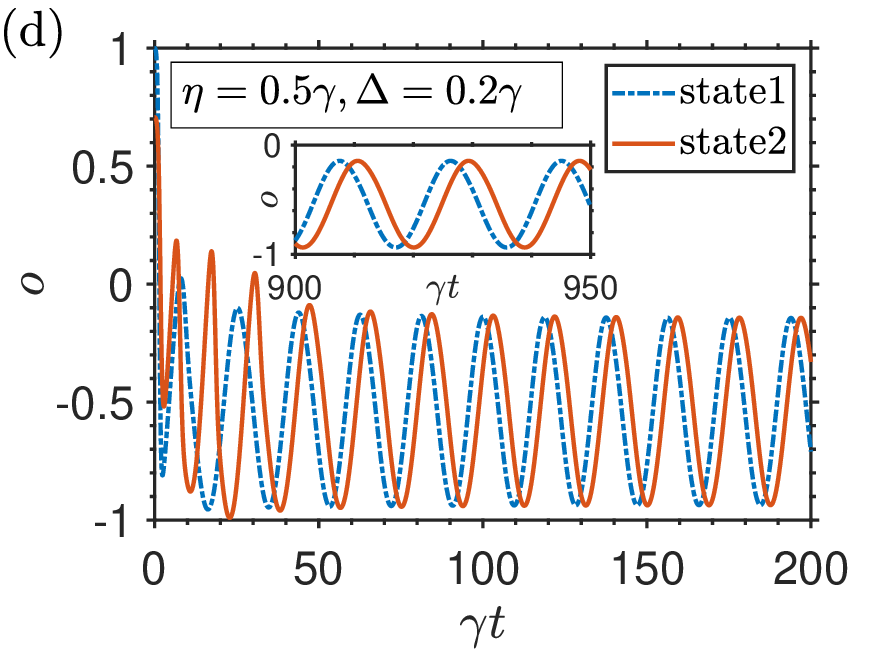}
  \caption{Mean-field dynamics of the order parameter for different initial states at representative BTC parameters with $\alpha=1$. In all panels, the two initial states are chosen as: state 1, $m_x^{(1,2)}=m_y^{(1,2)}=0$ and $m_z^{(1,2)}=1$; state 2, $m_x^{(1,2)}=1/\sqrt{2}$, $m_y^{(1,2)}=0$, and $m_z^{(1,2)}=1/\sqrt{2}$.}
  \label{contrast}
\end{figure}

For $\Delta=0$, the system exhibits neutral oscillations for arbitrary $\eta>0$, whose amplitudes depend on the initial state, as shown in Figs.~\ref{contrast}(a) and \ref{contrast}(b). Such initial-state-dependent oscillations are consistent with the neutral BTC dynamics reported in Ref.~\cite{FI2018}.

In contrast, for $\Delta\neq0$, trajectories starting from different initial states converge to the same periodic orbit and differ only by an overall phase shift, as shown in Figs.~\ref{contrast}(c) and \ref{contrast}(d). Their identical amplitudes and frequencies demonstrate the emergence of a unique attracting limit cycle selected by the system parameters.

\section{Analysis for two atomic ensembles}\label{B}

In this section, we first demonstrate the existence of a family of dissipation-free and low-dissipation modes. We then analyze the interaction structure on the effective triangular lattice and derive the corresponding semiclassical equations of motion in the thermodynamic limit. Finally, by examining the semiclassical trajectories in phase space, we show how these dissipation-free and low-dissipation modes provide protected dynamical channels that enable the BTC to persist under weak driving and finite detuning.

\subsection{Basis states of the composite Hilbert space}

Each atomic ensemble resides in the fully symmetric spin space
$\mathcal{H}_{S}^{(i)}$ with total spin quantum number $S=N_a/2$, which is spanned by the common eigenstates of
$({\bm S}^{(i)})^2=\sum_{\mu=x,y,z}(S_{\mu}^{(i)})^2$
and $S_z^{(i)}$. The Hilbert space of the composite system is therefore
$\mathcal{H}=\mathcal{H}_{S}^{(1)}\otimes\mathcal{H}_{S}^{(2)}$.
According to the Clebsch--Gordan (CG) decomposition~\cite{AC1866,HW1931,EP1959}, this product space can be decomposed into a direct sum of total-angular-momentum sectors,
\begin{align}
\mathcal{H}
=
\mathcal{H}_{S}^{(1)}\otimes\mathcal{H}_{S}^{(2)}
=
\bigoplus_{j=0}^{2S}\mathcal{T}_{j}.
\end{align}
Here, $\mathcal{T}_j$ denotes the sector with total spin quantum number $j$ and is spanned by the basis states
$\{\ket{j,m}\mid m=-j,-j+1,\ldots,j\}$, satisfying
${\bm J}^2\ket{j,m}=j(j+1)\ket{j,m},\,
J_z\ket{j,m}=m\ket{j,m}$,
where $J_\mu=S_\mu^{(1)}+S_\mu^{(2)}$ for $\mu=x,y,z$ and
${\bm J}^2=\sum_{\mu=x,y,z}J_\mu^2$.

To eliminate the phase $\phi$ from the collective jump operator, we introduce the unitary transformation
$U=\exp(i\phi S_z^{(2)})$ and define the rotated basis states as
\begin{align}
\ket{\varphi_{j,m}}\equiv U\ket{j,m}.
\end{align}
Since $U$ is unitary, the states
$\{\ket{\varphi_{j,m}}\mid j=0,1,\ldots,N_a,\ |m|\leq j\}$
form a complete orthonormal basis of the composite Hilbert space, satisfying
$\langle\varphi_{j',m'}|\varphi_{j,m}\rangle
=\delta_{j,j'}\delta_{m,m'},\,
\sum_{j=0}^{N_a}\sum_{m=-j}^{j}
\ket{\varphi_{j,m}}\bra{\varphi_{j,m}}
=\mathbb{I}$.

\subsection{Effective triangular lattices}

As discussed above, the rotated basis states
$\{\ket{\varphi_{j,m}}\mid j=0,1,\dots,N_a,\ |m|\leq j\}$
can be naturally arranged into the triangular basis lattice illustrated in Fig.~2(a) of the main text, with each lattice site representing an individual basis state. We now derive the effective inter-site processes generated by the master equation Eq.~(\ref{MES}), including both dissipative transitions and coherent couplings.

We first consider the incoherent transitions induced by the Lindblad dissipator. The population of the lattice site $(j,m)$ is defined as
$P_{j,m}=\mathrm{Tr}[\ket{\varphi_{j,m}}\bra{\varphi_{j,m}}\rho(t)]$.
Using $L=UJ_-U^\dagger$ and the angular-momentum relation
$J_{\pm}\ket{j,m}=\sqrt{(j\mp m)(j\pm m+1)}\,\ket{j,m\pm1}$,
we obtain
\begin{align}
\frac{dP_{j,m}}{dt}
&=\frac{\gamma}{2S}\Big[
2\langle\mathcal{L}^{\dagger}\ket{\varphi_{j,m}}\bra{\varphi_{j,m}}\mathcal{L}\rangle
-\langle\ket{\varphi_{j,m}}\bra{\varphi_{j,m}}\mathcal{L}^{\dagger}\mathcal{L}\rangle
-\langle\mathcal{L}^{\dagger}\mathcal{L}\ket{\varphi_{j,m}}\bra{\varphi_{j,m}}\rangle
\Big]
\nonumber\\
&=\frac{\gamma}{2S}\Big[
2\langle UJ_{+}\ket{j,m}\bra{j,m}J_{-}U^{\dagger}\rangle
-\langle U\ket{j,m}\bra{j,m}J_{+}J_{-}U^{\dagger}\rangle
-\langle UJ_{+}J_{-}\ket{j,m}\bra{j,m}U^{\dagger}\rangle
\Big]
\nonumber\\
&=\frac{\gamma}{S}\Big[
-(j+m)(j-m+1)P_{j,m}
+(j+m+1)(j-m)P_{j,m+1}
\Big]
\nonumber\\
&=-\Gamma_{j,m}P_{j,m}
+\Gamma_{j,m+1}P_{j,m+1},
\label{population_dissipation}
\end{align}
where the site-dependent dissipative transition rate is defined as
\begin{align}
\Gamma_{j,m}
=
\frac{\gamma}{S}(j+m)(j-m+1).
\label{Gamma_jm}
\end{align}
The Lindblad dissipation therefore generates a cascade of unidirectional transitions along each column of the triangular lattice, as indicated by the red arrows in Fig.~2(b) of the main text.

Second, we investigate the coherent inter-site couplings generated by the Hamiltonian
$H=\Delta J_z+\eta S_x^{(1)}$. According to the angular-momentum selection rules, the operator $S_x^{(1)}$ couples the basis states
$\ket{\varphi_{j,m}}\leftrightarrow\ket{\varphi_{j,m\pm1}}$
and
$\ket{\varphi_{j,m}}\leftrightarrow\ket{\varphi_{j\pm1,m\pm1}}$.
Using the completeness of the rotated basis, the Hamiltonian can be expressed as
\begin{align}
H
&=\sum_{j=0}^{N_a}\sum_{m=-j}^{j}
\sum_{j'=0}^{N_a}\sum_{m'=-j'}^{j'}
\ket{\varphi_{j',m'}}\bra{\varphi_{j',m'}}
\left(\Delta J_z+\eta S_x^{(1)}\right)
\ket{\varphi_{j,m}}\bra{\varphi_{j,m}}
\nonumber\\
&=\sum_{j=0}^{N_a}\sum_{m=-j}^{j}
\Delta m\,\ket{\varphi_{j,m}}\bra{\varphi_{j,m}}
+\eta\Bigg[
\sum_{j=0}^{N_a}\sum_{m=-j}^{j-1}
g_{j,m}^{\parallel}
\ket{\varphi_{j,m+1}}\bra{\varphi_{j,m}}
\nonumber\\
&\qquad+
\sum_{j=0}^{N_a-1}\sum_{m=-j}^{j}
\left(
g_{j,m}^{\nearrow}
\ket{\varphi_{j+1,m+1}}\bra{\varphi_{j,m}}
+
g_{j,m}^{\searrow}
\ket{\varphi_{j+1,m-1}}\bra{\varphi_{j,m}}
\right)
+\mathrm{H.c.}
\Bigg].
\label{H_two}
\end{align}
Here, $g_{j,m}^{\parallel}$, $g_{j,m}^{\nearrow}$, and
$g_{j,m}^{\searrow}$ denote the three types of coherent inter-site coupling shown in Fig.~2(b) of the main text. Since $[U,S_x^{(1)}]=0$, the coupling matrix elements can be evaluated directly in the unrotated basis $\{\ket{j,m}\}$. Therefore, their explicit expressions are
\begin{align}
g_{j,m}^{\parallel}
&=
\bra{j,m+1}S_x^{(1)}\ket{j,m}
=
\frac{1}{4}\sqrt{(j-m)(j+m+1)},
\label{g_parallel}\\
g_{j,m}^{\nearrow}
&=\bra{j+1,m+1}S_x^{(1)}\ket{j,m}
=
-\frac{1}{4}
\sqrt{
\frac{
(j+m+1)(j+m+2)(N_a-j)(N_a+j+2)
}{
(2j+1)(2j+3)
}
},
\label{g_up}\\
g_{j,m}^{\searrow}
&=\bra{j+1,m-1}S_x^{(1)}\ket{j,m}
=
\frac{1}{4}
\sqrt{
\frac{
(j-m+1)(j-m+2)(N_a-j)(N_a+j+2)
}{
(2j+1)(2j+3)
}
}.
\label{g_down}
\end{align}
The coupling $g_{j,m}^{\parallel}$ connects nearest-neighbor sites within the same column, whereas
$g_{j,m}^{\nearrow}$ and $g_{j,m}^{\searrow}$ connect sites in adjacent columns along the two diagonal directions, respectively. These three coherent couplings are represented by the corresponding double-headed arrows in Fig.~2(b) of the main text.

\subsection{Semiclassical equations of motion in the thermodynamic limit}

An arbitrary quantum state of the system can be expanded in the rotated basis as
\begin{align}
\ket{\psi(t)}
=
\sum_{j=0}^{N_a}\sum_{m=-j}^{j}
\psi_{j,m}(t)\ket{\varphi_{j,m}},
\end{align}
which can be interpreted as a quantum wave packet distributed over the triangular lattice. The coherent evolution of the system is therefore mapped onto the propagation of this wave packet among the lattice sites. Using the Schr\"odinger equation
$i\partial_t\ket{\psi(t)}=H\ket{\psi(t)}$, we obtain the equations of motion for the probability amplitudes,
\begin{align}
i\dot{\psi}_{j,m}(t)
&=
\Delta m\,\psi_{j,m}(t)
+\eta\Big[
g_{j,m}^{\parallel}\psi_{j,m+1}(t)
+g_{j,m-1}^{\parallel}\psi_{j,m-1}(t)
\nonumber\\
&\quad
+g_{j,m}^{\nearrow}\psi_{j+1,m+1}(t)
+g_{j-1,m-1}^{\nearrow}\psi_{j-1,m-1}(t)
\nonumber\\
&\quad
+g_{j,m}^{\searrow}\psi_{j+1,m-1}(t)
+g_{j-1,m+1}^{\searrow}\psi_{j-1,m+1}(t)
\Big].
\label{Sch}
\end{align}
To include all three types of inter-site couplings without boundary corrections, we restrict the following analysis to bulk lattice sites satisfying
$2\leq j\leq N_a-1$ and $|m|\leq j-2$.

In the thermodynamic limit $N_a\rightarrow\infty$, we introduce the continuous coordinates
$x=j/N_a$ and $z=m/N_a$. The bulk region then corresponds to
$0<x<1$ and $|z|<x$. Accordingly, the discrete wave packet can be represented in terms of the continuous coordinates as
\begin{align}
\ket{\psi(t)}
=
\iint \rho(x,z,t)\ket{\varphi_{x,z}}\,dx\,dz,
\end{align}
where $|\rho(x,z,t)|^2dx\,dz$ represents the occupation probability within the infinitesimal region
$[x,x+dx]\times[z,z+dz]$.
To derive the semiclassical equations of motion, we first express the inter-site coupling strengths in terms of $x$ and $z$ and expand them in powers of $1/N_a$.

For the coupling along the $m$ direction, we obtain
\begin{align}
g_{j,m}^{\parallel}
&=
\frac{N_a}{4}
\sqrt{(x-z)\left(x+z+\frac{1}{N_a}\right)}
\nonumber\\
&=
\frac{N_a}{4}
\left[
\sqrt{(x-z)(x+z)}
+
\frac{\sqrt{x-z}}{2\sqrt{x+z}}\frac{1}{N_a}
+
\mathcal{O}\!\left(\frac{1}{N_a^2}\right)
\right]
\nonumber\\
&=
\frac{N_a}{4}\sqrt{x^2-z^2}
+\mathcal{O}(1).
\label{g1}
\end{align}
Similarly,
\begin{align}
g_{j,m-1}^{\parallel}
&=
\frac{N_a}{4}
\sqrt{
\left(x-z+\frac{1}{N_a}\right)
\left(x+z-\frac{1}{N_a}\right)
}
\nonumber\\
&=
\frac{N_a}{4}
\left[
\sqrt{x^2-z^2}
+
\frac{z}{\sqrt{x^2-z^2}}\frac{1}{N_a}
+
\mathcal{O}\!\left(\frac{1}{N_a^2}\right)
\right]
\nonumber\\
&=
\frac{N_a}{4}\sqrt{x^2-z^2}
+\mathcal{O}(1).
\end{align}
The remaining coupling strengths can be expanded analogously, yielding
\begin{align}
g_{j,m}^{\nearrow}
&=g_{j-1,m-1}^{\nearrow}=
-\frac{N_a(x+z)\sqrt{1-x^2}}{8x}
+\mathcal{O}(1),\\
\\
g_{j,m}^{\searrow}
&=g_{j-1,m+1}^{\searrow}=
\frac{N_a(x-z)\sqrt{1-x^2}}{8x}
+\mathcal{O}(1).
\end{align}

For the probability amplitude $\psi_{j,m}(t)$, we employ a two-dimensional WKB ansatz
\begin{align}
\psi_{j,m}(t)
=
A(x,z,t)e^{iN_a I(x,z,t)},
\label{WKB}
\end{align}
where $A(x,z,t)$ denotes a slowly varying amplitude and $I(x,z,t)$ is the action. The canonical momenta conjugate to the continuous coordinates $x$ and $z$ are defined as
$p_x=\partial_x I(x,z,t)$ and $p_z=\partial_z I(x,z,t)$, respectively.

We first consider neighboring lattice sites within the same column. Using the WKB ansatz, we obtain
\begin{align}
\psi_{j,m\pm1}(t)
&=
A\left(x,z\pm\frac{1}{N_a},t\right)
\exp\left[
iN_a I\left(x,z\pm\frac{1}{N_a},t\right)
\right]
\nonumber\\
&=
\left[
A(x,z,t)+\mathcal{O}\left(\frac{1}{N_a}\right)
\right]
\exp\left\{
iN_a\left[
I(x,z,t)
\pm\frac{\partial_z I}{N_a}
+\mathcal{O}\left(\frac{1}{N_a^2}\right)
\right]
\right\}
\nonumber\\
&=
\psi_{j,m}(t)e^{\pm ip_z}
+\mathcal{O}\left(\frac{1}{N_a}\right).
\label{psi_parallel}
\end{align}
Similarly, for the two diagonal directions connecting adjacent columns, we have
\begin{align}
\psi_{j\pm1,m\pm1}(t)
&=
\psi_{j,m}(t)e^{\pm i(p_x+p_z)}
+\mathcal{O}\left(\frac{1}{N_a}\right),
\label{psi_nearrow}\\
\psi_{j\pm1,m\mp1}(t)
&=
\psi_{j,m}(t)e^{\pm i(p_x-p_z)}
+\mathcal{O}\left(\frac{1}{N_a}\right).
\label{psi_searrow}
\end{align}
These three relations correspond respectively to the lattice directions associated with the couplings
$g_{j,m}^{\parallel}$, $g_{j,m}^{\nearrow}$, and $g_{j,m}^{\searrow}$.

Substituting Eqs.~(\ref{g1})--(\ref{psi_searrow}) into Eq.~(\ref{Sch}) and retaining only the leading-order terms of $\mathcal{O}(N_a)$, we obtain
\begin{align}
\frac{\partial I}{\partial t}
+\Delta z
+\eta\Bigg[
&\frac{\sqrt{x^2-z^2}}{2}\cos p_z
-\frac{(x+z)\sqrt{1-x^2}}{4x}\cos(p_x+p_z)
+\frac{(x-z)\sqrt{1-x^2}}{4x}\cos(p_x-p_z)
\Bigg]
=0.
\label{HJ}
\end{align}
Equation~(\ref{HJ}) takes the standard Hamilton--Jacobi form
$\partial_t I+H_{\rm cl}=0$, where the effective semiclassical Hamiltonian governing the coherent dynamics in the thermodynamic limit is given by
\begin{align}
H_{\rm cl}(x,z,p_x,p_z)
=
\Delta z
+\eta\Bigg[
&\frac{\sqrt{x^2-z^2}}{2}\cos p_z
-\frac{(x+z)\sqrt{1-x^2}}{4x}\cos(p_x+p_z)
+\frac{(x-z)\sqrt{1-x^2}}{4x}\cos(p_x-p_z)
\Bigg].
\label{Hcl}
\end{align}

The coherent contribution to the semiclassical dynamics is governed by the canonical Hamilton equations
\begin{align}
\dot{x}_{\rm coh}
=\frac{\partial H_{\rm cl}}{\partial p_x},
\qquad
\dot{p}_{x,{\rm coh}}
=-\frac{\partial H_{\rm cl}}{\partial x},
\qquad
\dot{z}_{\rm coh}
=\frac{\partial H_{\rm cl}}{\partial p_z},
\qquad
\dot{p}_{z,{\rm coh}}
=-\frac{\partial H_{\rm cl}}{\partial z}.
\label{canonical_coh}
\end{align}

We next incorporate the correction induced by incoherent dissipation. As discussed above, the Lindblad dissipator generates unidirectional cascade transitions
$\ket{\varphi_{j,m}}\rightarrow\ket{\varphi_{j,m-1}}$
along each column of the triangular lattice. Since such a transition leaves $j$ unchanged while decreasing $m$ by one, dissipation produces a drift along the $z$ direction without directly modifying $x$.

Within the quantum-jump picture, during an infinitesimal time interval $dt$, the transition
$\ket{\varphi_{j,m}}\rightarrow\ket{\varphi_{j,m-1}}$
occurs with probability $\Gamma_{j,m}dt$. Since each jump changes the continuous coordinate $z=m/N_a$ by $-1/N_a$, the average dissipative drift is given by
\begin{align}
dz_{\rm diss}
&=
-\frac{1}{N_a}\Gamma_{j,m}\,dt
\nonumber\\
&=
-\frac{1}{N_a}
\frac{\gamma}{S}(j+m)(j-m+1)\,dt
\nonumber\\
&=
-2\gamma(x+z)
\left(
x-z+\frac{1}{N_a}
\right)dt.
\end{align}
In the thermodynamic limit $N_a\rightarrow\infty$, the subleading terms of order $\mathcal{O}(1/N_a)$ can be neglected, yielding
\begin{align}
\dot{z}_{\rm diss}
=
-2\gamma(x^2-z^2).
\label{zdiss}
\end{align}
The fluctuations associated with individual quantum jumps give rise to higher-order corrections that vanish in the thermodynamic limit, leaving the deterministic dissipative drift in Eq.~(\ref{zdiss}).

Combining the coherent Hamiltonian dynamics with the dissipative drift, we obtain the semiclassical equations of motion in the thermodynamic limit,
\begin{align}
\dot{x}
&=
-\frac{\eta\sqrt{1-x^2}}{4x}
\left[
(x-z)\sin(p_x-p_z)
-(x+z)\sin(p_x+p_z)
\right],
\nonumber\\
\dot{z}
&=
-\frac{\eta}{2}\sqrt{x^2-z^2}\sin p_z
+\frac{\eta\sqrt{1-x^2}}{4x}
\left[
(x-z)\sin(p_x-p_z)
+(x+z)\sin(p_x+p_z)
\right]
\nonumber\\
&\quad
-2\gamma(x^2-z^2),
\nonumber\\
\dot{p}_x
&=
-\frac{\eta x}{2\sqrt{x^2-z^2}}\cos p_z
+\frac{\eta}{4x^2\sqrt{1-x^2}}
\left[
(x-z)\cos(p_x-p_z)
-(x+z)\cos(p_x+p_z)
\right]
\nonumber\\
&\quad
-\frac{\eta\sqrt{1-x^2}}{4x}
\left[
\cos(p_x-p_z)-\cos(p_x+p_z)
\right],
\nonumber\\
\dot{p}_z
&=
-\Delta
+\frac{\eta z}{2\sqrt{x^2-z^2}}\cos p_z
+\frac{\eta\sqrt{1-x^2}}{4x}
\left[
\cos(p_x-p_z)+\cos(p_x+p_z)
\right].
\label{MOE_xz}
\end{align}
As shown below, the semiclassical trajectories obtained from Eq.~(\ref{MOE_xz}) coincide with the corresponding mean-field dynamics, confirming the validity of the dissipative correction in the thermodynamic limit.

\subsection{Semiclassical phase-space trajectories}

Within the narrow-wave-packet approximation,
\begin{align}
|\rho(x,z,t)|^2
\simeq
\delta\big[x-x(t)\big]\delta\big[z-z(t)\big],
\end{align}
the order parameter can be expressed as
\begin{align}
o(t)
&=
\frac{\langle J_z\rangle}{N_a}
=
\frac{\bra{\psi(t)}J_z\ket{\psi(t)}}{N_a}
=
\iint |\rho(x,z,t)|^2 z\,dx\,dz
=
z(t).
\label{order_parameter_z}
\end{align}
Therefore, the evolution of the wave-packet center $z(t)$ governed by Eq.~(\ref{MOE_xz}) directly reproduces the dynamics of the order parameter $o(t)$.

\begin{figure}
\centering
\includegraphics[width=0.48\columnwidth]{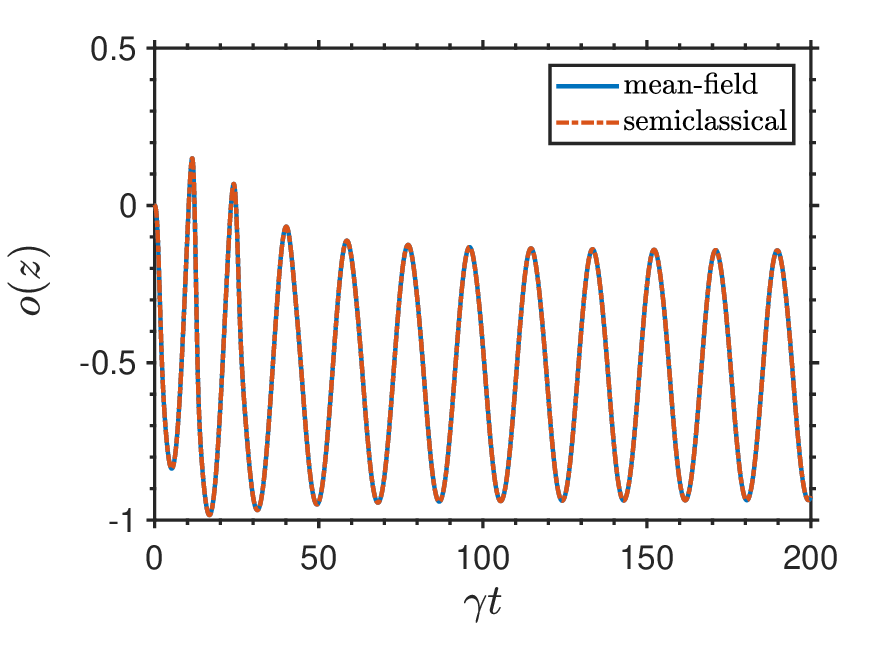}
\caption{Comparison between the order-parameter dynamics obtained from the mean-field equations and the semiclassical equations of motion. For the mean-field calculation, the initial conditions are $m_{x,y}^{(1,2)}(0)=0$, $m_z^{(1)}(0)=1$, and $m_z^{(2)}(0)=-1$. For the semiclassical calculation, we take $x(0)=10^{-7}$ and $z(0)=p_x(0)=p_z(0)=0$. The parameters are $\eta=0.5\gamma$, $\Delta=0.2\gamma$.}
\label{traj}
\end{figure}

To verify the semiclassical treatment, we consider the initial state
$\ket{\psi(0)}=\ket{S,S}\otimes\ket{S,-S}$,
which corresponds to the mean-field initial conditions
$m_{x,y}^{(1,2)}(0)=0$,
$m_z^{(1)}(0)=1$, and
$m_z^{(2)}(0)=-1$.
The corresponding evolution of the order parameter obtained from the mean-field equations Eq.~(\ref{MF}) is shown by the solid curve in Fig.~\ref{traj}.
In the total-angular-momentum basis, this initial state can be expanded as
$\ket{\psi(0)}=\sum_{j=0}^{N_a}\psi_{j,0}(0)\ket{j,0}$,
$\psi_{j,0}(0)
=\sqrt{2j+1}\,N_a!/\sqrt{(N_a-j)!(N_a+j+1)!}$.

The distribution $|\psi_{j,0}(0)|^2$ has a width
$\delta j\sim\sqrt{N_a}$ and $\delta m=0$, with its center located near
$j_c=\left\lfloor\sqrt{\frac{N_a+1}{2}}\right\rfloor,
\,m_c=0$, where $\lfloor\cdot\rfloor$ denotes the floor function.
In terms of the continuous coordinates, the corresponding macroscopic widths satisfy
$\delta x=\lim_{N_a\rightarrow\infty}(\delta j/N_a)=0,\,\delta z=0$,
while the center approaches $x_c=
\lim_{N_a\rightarrow\infty}(j_c/N_a)=0,\,z_c=0$.

Thus, the initial state corresponds to a macroscopically narrow wave packet in the thermodynamic limit. The evolution of $z(t)$ obtained from Eq.~(\ref{MOE_xz}) is shown by the dot-dashed curve in Fig.~\ref{traj}. Since the semiclassical equations contain terms that are singular exactly at $x=0$, we take a small positive value $x(0)=10^{-7}$ in the numerical calculation. The excellent agreement between the semiclassical and mean-field results confirms the validity of the semiclassical treatment and the dissipative correction introduced above.

To reveal the role of the low-dissipation modes in sustaining the BTC, we introduce the rotated canonical coordinates
\begin{align}
v=\frac{x+z}{\sqrt{2}},
\qquad
u=\frac{x-z}{\sqrt{2}},
\end{align}
where $v$ and $u$ measure the perpendicular distances from the lower and upper boundaries of the triangular lattice, respectively. The corresponding conjugate momenta are
\begin{align}
p_v=\frac{p_x+p_z}{\sqrt{2}},
\qquad
p_u=\frac{p_x-p_z}{\sqrt{2}}.
\end{align}
In terms of these variables, Eq.~(\ref{MOE_xz}) becomes
\begin{align}
\dot{v}
&=
\frac{\eta B}{\sqrt{2}(u+v)}
u\sin(\sqrt{2}p_u)
-\frac{\eta}{2}\sqrt{uv}\sin\delta
-2\sqrt{2}\gamma uv,
\nonumber\\
\dot{u}
&=
-\frac{\eta B}{\sqrt{2}(u+v)}
v\sin(\sqrt{2}p_v)
+\frac{\eta}{2}\sqrt{uv}\sin\delta
+2\sqrt{2}\gamma uv,
\nonumber\\
\dot{p}_v
&=
-\frac{\Delta}{\sqrt{2}}
-\frac{\eta}{2\sqrt{2}}
\sqrt{\frac{u}{v}}\cos\delta
+\frac{\eta}{2(u+v)^2B}
\Big[
u\cos(\sqrt{2}p_u)
-v\cos(\sqrt{2}p_v)
\Big]+\frac{\eta B}{2(u+v)}
\cos(\sqrt{2}p_v),
\nonumber\\
\dot{p}_u
&=
\frac{\Delta}{\sqrt{2}}
-\frac{\eta}{2\sqrt{2}}
\sqrt{\frac{v}{u}}\cos\delta
+\frac{\eta}{2(u+v)^2B}
\Big[
u\cos(\sqrt{2}p_u)
-v\cos(\sqrt{2}p_v)
\Big]-\frac{\eta B}{2(u+v)}
\cos(\sqrt{2}p_u),
\label{MOE_uv}
\end{align}
where
\begin{align}
B=\sqrt{1-\frac{(u+v)^2}{2}},
\qquad
\delta=\frac{p_v-p_u}{\sqrt{2}}.
\end{align}

Figure~3(a) of the main text shows the long-time trajectories obtained from Eq.~(\ref{MOE_uv}), projected onto the $u$--$v$ plane for representative parameter points in the BTC phase.

\section{Semiclassical picture for a single atomic ensemble}\label{C}

In this section, we apply the same semiclassical analysis to the single-atomic-ensemble BTC introduced in Ref.~\cite{FI2018}, in order to elucidate the physical origin of its requirements of strong driving and exact resonance.

The dynamics is governed by the master equation
\begin{align}
\frac{d\rho}{dt}
=
-i[\Delta S_z+\eta S_x,\rho]
+\frac{\gamma}{2S}
\left(
2S_-\rho S_+
-S_+S_-\rho
-\rho S_+S_-
\right),
\label{ME_one}
\end{align}
which corresponds to Eq.~(1) of the main text with $\alpha=0$, where the ensemble superscript has been omitted for notational simplicity. Here, $\Delta=\omega_a-\omega_p$ denotes the detuning between the driving and atomic transition frequencies, $\eta$ is the driving strength, $\gamma$ is the collective dissipation strength, and $S=N_a/2$ denotes the collective spin quantum number.

The system resides in the collective-spin Hilbert space $\mathcal{H}_S$, spanned by the basis states $\{\ket{S,m}\mid -S\leq m\leq S\}$. These basis states can be arranged into the effective one-dimensional lattice illustrated in Fig.~2(c) of the main text. We first examine the coherent inter-site couplings generated by the Hamiltonian $H=\Delta S_z+\eta S_x$. Inserting the completeness relation gives
\begin{align}
H
&=
\sum_{m_1=-S}^{S}
\sum_{m_2=-S}^{S}
\ket{S,m_1}\bra{S,m_1}
\left(\Delta S_z+\eta S_x\right)
\ket{S,m_2}\bra{S,m_2}
\nonumber\\
&=
\sum_{m=-S}^{S}
\Delta m\ket{S,m}\bra{S,m}
+\frac{\eta}{2}
\sum_{m=-S}^{S-1}
t_m
\left(
\ket{S,m+1}\bra{S,m}
+\ket{S,m}\bra{S,m+1}
\right),
\label{H_one}
\end{align}
where
\begin{align}
t_m=\sqrt{(S-m)(S+m+1)}.
\end{align}
Thus, each lattice site $\ket{S,m}$ has an on-site energy $\Delta m$, while the driving term induces position-dependent nearest-neighbor coherent hopping with strength $\eta t_m/2$.

We next derive the semiclassical description of Eq.~(\ref{H_one}) in the thermodynamic limit. An arbitrary state in $\mathcal{H}_S$ can be expanded as
\begin{align}
\ket{\psi(t)}
=
\sum_{m=-S}^{S}\psi_m(t)\ket{S,m}.
\end{align}
The Schr\"odinger equation then yields the discrete equation for the probability amplitudes,
\begin{align}
i\dot{\psi}_m(t)
=
\Delta m\,\psi_m(t)
+\frac{\eta}{2}
\left[
t_m\psi_{m+1}(t)
+t_{m-1}\psi_{m-1}(t)
\right].
\label{Sch_one}
\end{align}

In the thermodynamic limit, we introduce the continuous coordinate $x=m/S$. For bulk lattice sites with $-1<x<1$, the hopping amplitudes satisfy
$t_m=t_{m-1}=S\sqrt{1-x^2}+\mathcal{O}(1)$,

We further employ the one-dimensional WKB ansatz
\begin{align}
\psi_m(t)
=
A(x,t)e^{iSI(x,t)},
\label{WKB_one}
\end{align}
where $A(x,t)$ is a slowly varying amplitude, $I(x,t)$ denotes the action, and the conjugate momentum is defined as $p=\partial_x I$. To leading order in $1/S$, the amplitudes at neighboring sites satisfy
\begin{align}
\psi_{m\pm1}(t)
=
\psi_m(t)e^{\pm ip}
+\mathcal{O}(1/S).
\end{align}
Substituting these expressions into Eq.~(\ref{Sch_one}) and retaining the leading-order terms of $\mathcal{O}(S)$, we obtain the Hamilton--Jacobi equation
\begin{align}
\frac{\partial I}{\partial t}
+
\Delta x
+
\eta\sqrt{1-x^2}\cos p
=
0.
\end{align}
The corresponding effective semiclassical Hamiltonian for the coherent dynamics is therefore
\begin{align}
H_{\rm cl}(x,p)
=
\Delta x
+\eta\sqrt{1-x^2}\cos p.
\label{Hcl_one}
\end{align}
The coherent contribution to the semiclassical equations of motion follows from the canonical Hamilton equations,
\begin{align}
\dot{x}_{\rm coh}
&=
\frac{\partial H_{\rm cl}}{\partial p}
=
-\eta\sqrt{1-x^2}\sin p,
\nonumber\\
\dot{p}_{\rm coh}
&=
-\frac{\partial H_{\rm cl}}{\partial x}
=
-\Delta
+\frac{\eta x}{\sqrt{1-x^2}}\cos p.
\label{canonical}
\end{align}

Next, we incorporate the dissipative contribution into the semiclassical equations of motion. The collective dissipation induces cascaded transitions
$\ket{S,m}\rightarrow\ket{S,m-1}$ along the effective one-dimensional lattice. The population
$P_m=\mathrm{Tr}\!\left[\ket{S,m}\bra{S,m}\rho\right]$
obeys
\begin{align}
\frac{dP_m}{dt}
=
-\Gamma_m P_m
+\Gamma_{m+1}P_{m+1},
\end{align}
where
\begin{align}
\Gamma_m
=
\frac{\gamma}{S}(S+m)(S-m+1)
\end{align}
is the transition rate from $\ket{S,m}$ to $\ket{S,m-1}$.

Within the quantum-jump picture, each dissipative transition changes the continuous coordinate $x=m/S$ by
$\delta x=-1/S$, while the probability for such a transition during an infinitesimal time interval $\mathrm{d}t$ is $\Gamma_m\mathrm{d}t$. The resulting dissipative drift is therefore
\begin{align}
\mathrm{d}x_{\rm diss}
=-\frac{\Gamma_m}{S}\,\mathrm{d}t
=-\gamma(1+x)
\left(1-x+\frac{1}{S}\right)\mathrm{d}t.
\end{align}
Taking the thermodynamic limit $S\rightarrow\infty$ gives
\begin{align}
\dot{x}_{\rm diss}
=
-\gamma(1-x^2).
\end{align}
The dissipation thus generates a deterministic drift toward the lower boundary $x=-1$, with a rate that vanishes at the boundary itself.

Combining this dissipative drift with the coherent dynamics in Eq.~(\ref{canonical}), we obtain the semiclassical equations of motion
\begin{align}
\dot{x}
&=
-\eta\sqrt{1-x^2}\sin p
-\gamma(1-x^2),
\nonumber\\
\dot{p}
&=
-\Delta
+\frac{\eta x}{\sqrt{1-x^2}}\cos p.
\label{MOE_one}
\end{align}

\section{Liouvillian spectrum and quantum analysis in the BTC phase}
\label{D}

In the main text, we characterize the quantum signatures of the BTC at a representative parameter point with weak driving and finite detuning. For clarity, only one representative branch of the low-lying Liouvillian spectrum is displayed in Fig.~4(b) of the main text. In this section, we provide two complementary analyses. First, we present the more complete low-lying Liouvillian spectrum to reveal its multi-branch structure. Second, we perform an analogous quantum analysis at weak driving and zero detuning, thereby complementing the finite-detuning results presented in the main text.

\begin{figure}
\centering
 \includegraphics[width=0.48\columnwidth]{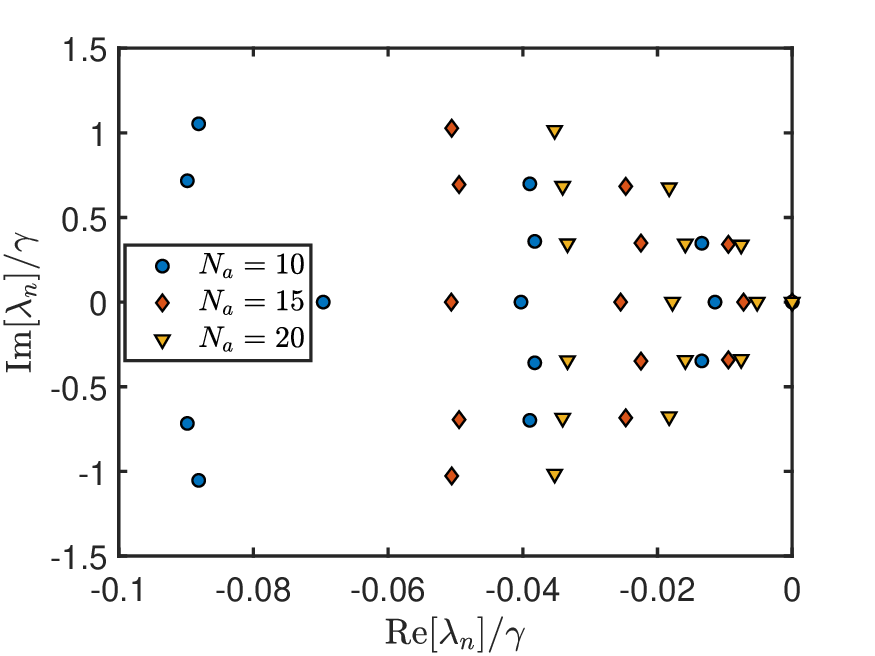}
  \caption{Low-lying Liouvillian spectrum for $N_a=10$, $15$, and $20$. The paraments are set as $\eta=0.5\gamma$, $\Delta=0.2\gamma$.}
  \label{spectrum}
\end{figure}

The Liouvillian eigenmodes are defined by
$\mathcal{L}(\rho_n)=\lambda_n\rho_n$,
where $\mathcal{L}$ denotes the Liouvillian superoperator associated with the master equation Eq.~(\ref{MES}), such that
$\dot{\rho}=\mathcal{L}(\rho)$.
Figure~\ref{spectrum} shows the low-lying Liouvillian spectra for
$N_a=10$, $15$, and $20$.
For each system size, the eigenvalues organize into multiple well-defined branches. Within each branch, the imaginary parts are approximately equally spaced, while the real parts approach zero as $N_a$ increases. The more complete low-lying spectrum therefore reveals that the spectral structure discussed in the main text is not restricted to the representative branch shown in Fig.~4(b), but extends to a family of low-lying oscillatory Liouvillian modes.

\begin{figure}
 \includegraphics[width=0.33\columnwidth]{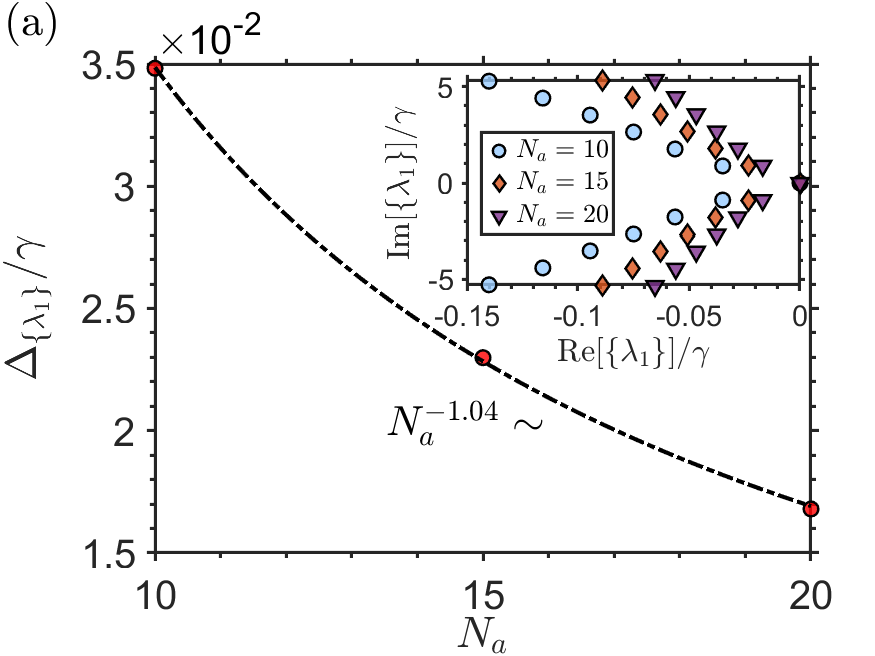}\includegraphics[width=0.33\columnwidth]{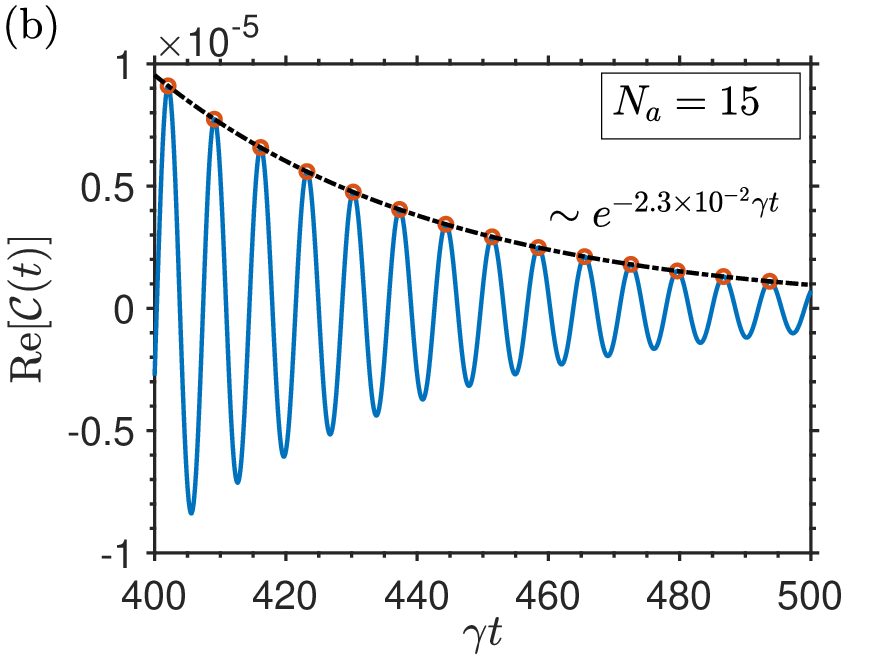}\includegraphics[width=0.33\columnwidth]{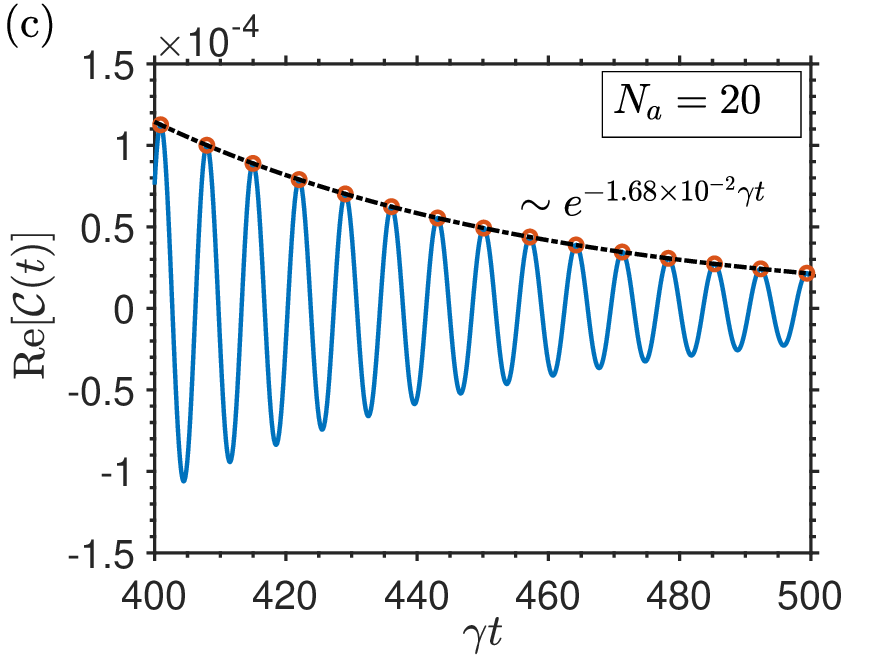}
  \caption{(a) Low-lying Liouvillian spectrum for the first eigenvalue branch at $N_a=10,\,15,\,20$. Inset in panel (a): power-law fitting of the first-branch energy gap versus $N_a$. (b) and (c) Long-time dynamics of the two-time correlation function of the order-parameter fluctuation operator for $N_a=15$ and $N_a=20$, respectively. The paraments are set as $\eta=0.9\gamma$, $\Delta=0$.}
  \label{quantum}
\end{figure}

We next complement the finite-detuning analysis of the main text by considering a representative BTC point at weak driving and exact resonance, with $\eta=0.9\gamma$ and $\Delta=0$. The inset in Fig.~\ref{quantum}(a) shows the first branch of the low-lying Liouvillian spectrum for different system sizes. The imaginary parts of neighboring eigenvalues exhibit an approximately uniform spacing that remains essentially unchanged with increasing $N_a$, whereas their real parts progressively approach the imaginary axis. As shown in the main figure, the Liouvillian gap associated with this branch decreases algebraically with the system size. These results indicate the emergence, in the thermodynamic limit, of a family of dissipation-free oscillatory modes with finite and regularly spaced frequencies.

Figures~\ref{quantum}(b) and \ref{quantum}(c) show the long-time evolution of $\mathrm{Re}[\mathcal{C}(t)]$ for $N_a=15$ and $20$, respectively. For finite $N_a$, the correlation function exhibits oscillations with an exponentially decaying envelope. The corresponding decay rate is consistent with the Liouvillian gap of the first spectral branch and decreases as the system size increases. Consequently, the lifetime of the order-parameter correlations grows with $N_a$ and diverges as the Liouvillian gap closes in the thermodynamic limit.

The coexistence of a closing Liouvillian gap, regularly spaced finite-frequency modes, and increasingly long-lived oscillations of the order-parameter correlation function provides complementary quantum evidence for the BTC at weak driving and zero detuning. Together with the finite-detuning results presented in the main text, these results demonstrate that the BTC persists throughout both resonant and finite-detuning regimes under weak driving.

\section{Implementation in a waveguide-QED setup}
\label{E}

In this section, we show that the collective dissipative dynamics considered in the main text can be naturally realized in a waveguide-QED platform. We consider two spatially separated ensembles of two-level atoms collectively coupled to a common one-dimensional waveguide. The total Hamiltonian is
\begin{align}
H
=
\int dk\,v|k|a_k^\dagger a_k
+\omega_a\left(S_z^{(1)}+S_z^{(2)}\right)
+\frac{g}{\sqrt{2\pi S}}
\int dk\,
\Big[
S_+^{(1)}a_k e^{-ikx_1}
+
e^{i\phi}S_+^{(2)}a_k e^{-ikx_2}
+\mathrm{H.c.}
\Big],
\label{H_waveguide}
\end{align}
where $v$ is the group velocity of the waveguide photons and
$\omega_a$ is the atomic transition frequency. For the $i$th ensemble,
$S_\mu^{(i)}$ ($\mu=x,y,z$) denotes the collective spin operator with
spin quantum number $S=N_a/2$, and
$S_\pm^{(i)}=S_x^{(i)}\pm iS_y^{(i)}$.
The two ensembles are located at $x_1$ and $x_2$ and couple to the
waveguide with the same strength $g$. The phase $\phi$ represents a
controllable relative phase between their couplings to the common
waveguide.

Taking
\begin{align}
H_0
=
\int dk\,v|k|a_k^\dagger a_k
+
\omega_a\left(S_z^{(1)}+S_z^{(2)}\right),
\end{align}
the atom--waveguide interaction in the interaction picture reads
\begin{align}
H_I(t)
=
\frac{g}{\sqrt{2\pi S}}
\int dk\,
\Big[
&S_+^{(1)}a_k e^{-ikx_1}
+
e^{i\phi}S_+^{(2)}a_k e^{-ikx_2}
\Big]
e^{-i(v|k|-\omega_a)t}
+\mathrm{H.c.}
\label{HI}
\end{align}

Within the Born--Markov approximation, the reduced density matrix of
the two atomic ensembles obeys
\begin{align}
\frac{d\rho}{dt}
=
-\int_0^\infty d\tau\,
\mathrm{Tr}_w
\left[
H_I(t),
\left[
H_I(t-\tau),
\rho_w\otimes\rho(t)
\right]
\right],
\label{MER}
\end{align}
where $\mathrm{Tr}_w$ denotes the trace over the waveguide degrees of
freedom. We assume that the waveguide is initially in the vacuum state,
$\rho_w=\ket{0}\bra{0}$, substituting Eq.~(\ref{HI}) into Eq.~(\ref{MER}) gives
\begin{align}
\frac{d\rho}{dt}
=
\frac{g^2}{2\pi S}
\int dk\int_0^\infty d\tau\,
\Big\{
&e^{i(v|k|-\omega_a)\tau}
\Big[
S_-^{(1)}\rho S_+^{(1)}
+
e^{-ik\Delta x}e^{i\phi}
S_-^{(1)}\rho S_+^{(2)}
+e^{ik\Delta x}e^{-i\phi}
S_-^{(2)}\rho S_+^{(1)}
+
S_-^{(2)}\rho S_+^{(2)}
\Big]
\nonumber\\
-&e^{-i(v|k|-\omega_a)\tau}
\Big[
S_+^{(1)}S_-^{(1)}\rho
+
e^{ik\Delta x}e^{-i\phi}
S_+^{(1)}S_-^{(2)}\rho+
e^{-ik\Delta x}e^{i\phi}
S_+^{(2)}S_-^{(1)}\rho
+
S_+^{(2)}S_-^{(2)}\rho
\Big]
+\mathrm{H.c.}
\Big\},
\label{ME_waveguide_intermediate}
\end{align}
where $\Delta x=x_2-x_1$ is the ensemble separation.

Using
\begin{align}
\int_0^\infty d\tau\,e^{\pm i\Omega\tau}
=
\pi\delta(\Omega)
\pm i\,\mathcal{P}\frac{1}{\Omega},
\end{align}
the resonant contribution is selected at
$|k|=k_a=\omega_a/v$. The principal-value contribution produces a
Lamb-shift Hamiltonian, which can be absorbed into a renormalization of
the atomic transition frequency and is omitted here. We further choose
the ensemble separation such that
$k_a\Delta x=2\pi n,\, n\in\mathbb{Z}$,
or equivalently $\Delta x=2\pi nv/\omega_a$. The propagation phase
between the two ensembles is then unity at the atomic transition
frequency. Under these conditions, the reduced atomic dynamics becomes
\begin{align}
\frac{d\rho}{dt}
=
\frac{\gamma}{2S}
\Bigg\{
&\sum_{i=1,2}
\left(
2S_-^{(i)}\rho S_+^{(i)}
-
S_+^{(i)}S_-^{(i)}\rho
-
\rho S_+^{(i)}S_-^{(i)}
\right)
+
e^{i\phi}
\left(
2S_-^{(1)}\rho S_+^{(2)}
-
S_+^{(2)}S_-^{(1)}\rho
-
\rho S_+^{(2)}S_-^{(1)}
\right)
\nonumber\\
&+
e^{-i\phi}
\left(
2S_-^{(2)}\rho S_+^{(1)}
-
S_+^{(1)}S_-^{(2)}\rho
-
\rho S_+^{(1)}S_-^{(2)}
\right)
\Bigg\}=
\frac{\gamma}{2S}
\left(
2L\rho L^\dagger-L^\dagger L\rho-\rho L^\dagger L
\right),
\label{MEI}
\end{align}
where $L=S_-^{(1)}+e^{-i\phi}S_-^{(2)}$.
is the collective jump operator and $\gamma$ denotes the effective
waveguide-induced collective dissipation strength. Equation~(\ref{MEI})
therefore explicitly shows that coupling the two ensembles to the same
waveguide generates both the individual and correlated dissipative
processes required in our model.

We finally introduce a coherent drive with frequency $\omega_p$ and
strength $\eta$ applied to the first atomic ensemble. In the
Schr\"odinger picture, the atomic Hamiltonian is
\begin{align}
H_{\rm at}(t)
=
\omega_a
\left(
S_z^{(1)}+S_z^{(2)}
\right)
+
\frac{\eta}{2}
\left(
S_-^{(1)}e^{i\omega_p t}
+
S_+^{(1)}e^{-i\omega_p t}
\right).
\end{align}
Transforming to the frame rotating at the driving frequency through
\begin{align}
U_p(t)
=
\exp\left[
-i\omega_p
\left(
S_z^{(1)}+S_z^{(2)}
\right)t
\right],
\end{align}
we obtain
\begin{align}
\frac{d\rho}{dt}
=\mathcal{L}\rho=
-i
\left[
\Delta
\left(
S_z^{(1)}+S_z^{(2)}
\right)
+
\eta S_x^{(1)},
\rho
\right]
+
\frac{\gamma}{2S}
\left(
2L\rho L^\dagger
-L^\dagger L\rho
-\rho L^\dagger L\right),
\label{ME_waveguide_final}
\end{align}
where $\Delta=\omega_a-\omega_p$. Equation~(\ref{ME_waveguide_final})
is identical to Eq.~(\ref{MES}) and Eq.~(1) in the main text with $\alpha=1$, demonstrating that the dissipative
BTC model considered can be implemented within a
waveguide-QED architecture.

\begingroup

\renewcommand{\bibnumfmt}[1]{[S#1]}

\endgroup

\begin{thebibliography}{99}
%%% Symmetry breaking
\bibitem{FS2005} F. Strocchi, \textit{Symmetry Breaking}, Lect. Notes Phys. \textbf{643} (Springer, Berlin, 2005).
%%% TC_origin
\bibitem{FW2012} F. Wilczek, Quantum time crystals, Phys. Rev. Lett. {\bf 109}, 160401 (2012).
%%% no-go
\bibitem{PB2013} P. Bruno, Impossibility of Spontaneously Rotating Time Crystals: A No-Go Theorem, Phys. Rev. Lett. {\bf 111}, 070402 (2013).
\bibitem{HW2015} H. Watanabe, and M. Oshikawa, Absence of Quantum Time Crystals, Phys. Rev. Lett. {\bf 114}, 251603 (2015).
%%% DTC
\bibitem{DV2016} D. V. Else, B. Bauer, and C. Nayak, Floquet Time Crystals, Phys. Rev. Lett. {\bf 117}, 090402 (2016).
\bibitem{VK2016} V. Khemani, A. Lazarides, R. Moessner, and S. L. Sondhi, Phase Structure of Driven Quantum Systems, Phys. Rev. Lett. {\bf 116}, 250401 (2016).
\bibitem{SC2017}S. Choi, J. Choi, R. Landig, G. Kucsko, H. Zhou, J. Isoya, F. Jelezko, S. Onoda, H. Sumiya, V. Khemani, C. von Keyserlingk, N. Y. Yao, E. Demler, and M. D. Lukin, Observation of discrete time-crystalline order in a disordered dipolar many-body system, Nature {\bf 543}, 221 (2017).
\bibitem{JZ2017}J. Zhang, P. W. Hess, A. Kyprianidis, P. Becker, A. Lee, J. Smith, G. Pagano, I.-D. Potirniche, A. C. Potter, A. Vishwanath, N. Y. Yao, and C. Monroe, Observation of a discrete time crystal, Nature, {\bf 543}, 217 (2017).
\bibitem{NY2017} N. Y. Yao, A. C. Potter, I.-D. Potirniche, and A. Vishwanath, Discrete Time Crystals: Rigidity, Criticality, and Realizations, Phys. Rev. Lett. {\bf 118}, 030401 (2017).
\bibitem{DV2017} D. V. Else, B. Bauer, and C. Nayak, Prethermal Phases of Matter Protected by Time-Translation Symmetry, Phys. Rev. X {\bf 7}, 011026 (2017).
\bibitem{WW2017} W. W. Ho, S. Choi, M. D. Lukin, and D. A. Abanin, Critical Time Crystals in Dipolar Systems, Phys. Rev. Lett. {\bf 119}, 010602 (2017).
\bibitem{ZG2018} Z. Gong, R. Hamazaki, and M. Ueda, Discrete Time-Crystalline Order in Cavity and Circuit QED Systems, Phys. Rev. Lett. {\bf 120}, 040404 (2018).
\bibitem{SP2018} S. Pal, N. Nishad, T. S. Mahesh, and G. J. Sreejith, Temporal Order in Periodically Driven Spins in Star-Shaped Clusters, Phys. Rev. Lett. {\bf 120}, 180602 (2018).
\bibitem{JR2018} J. Rovny, R. L. Blum, and S. E. Barrett, Observation of Discrete-Time-Crystal Signatures in an Ordered Dipolar Many-Body System, Phys. Rev. Lett. {\bf 120}, 180603 (2018).
\bibitem{DV2020} D. V. Else, C. Monroe, C. Nayak, and N. Y. Yao, Discrete Time Crystals, Annu. Rev. Condens. Matter Phys. {\bf 11}, 467 (2020).
\bibitem{NY2020} N. Y. Yao, C. Nayak, L. Balents, and M. P. Zaletel, Classical Discrete Time Crystals, Nat. Phys. {\bf 16}, 438 (2020).
\bibitem{AK2021} A. Kyprianidis, F. Machado, W. Morong, P. Becker, K. S. Collins, D. V. Else, L. Feng, P. W. Hess, C. Nayak, G. Pagano, N. Y. Yao, and C. Monroe, Observation of a prethermal discrete time crystal, Science {\bf 372}, 1192 (2021).
\bibitem{JR2021} J. Randall, C. E. Bradley, F. V. van der Gronden, A. Galicia, M. H. Abobeih, M. Markham, D. J. Twitchen, F. Machado, N. Y. Yao, and T. H. Taminiau, Many-body-localized discrete time crystal with a programmable spin-based quantum simulator, Science {\bf 374}, 1474 (2021).
\bibitem{XM2022} X. Mi et al., Time-crystalline eigenstate order on a quantum processor, Nature {\bf 601}, 531 (2022).
%%% CTC_dissipate
\bibitem{KS2018} K. Sacha, and J. Zakrzewski, Time Crystals: A Review, Rep. Prog. Phys. {\bf 81}, 016401 (2018).
\bibitem{FI2018x} F. Iemini, A. Russomanno, J. Keeling, M. Schir\`{o}, M. Dalmonte, and R. Fazio, Boundary Time Crystals, Phys. Rev. Lett. {\bf 121}, 035301 (2018).
\bibitem{SA2018} S. Autti, V. B. Eltsov, and G. E. Volovik, Observation of a Time Quasicrystal and Its Transition to a Superfluid Time Crystal, Phys. Rev. Lett. {\bf 120}, 215301 (2018).
\bibitem{BB2019} B. Bu\v{c}a, J. Tindall, and D. Jaksch, Non-stationary coherent quantum many-body dynamics through dissipation. Nat Commun {\bf 10}, 1730 (2019).
\bibitem{AR2020} A. Riera-Campeny, M. Moreno-Cardoner, and A. Sanpera, Time Crystallinity in Open Quantum Systems, Quantum {\bf 4}, 270 (2020).
\bibitem{LF2021} L. F. d. Prazeres, L. d. S. Souza, and F. Iemini, Boundary time crystals in collective d-level systems, Phys. Rev. B, {\bf 103},184308 (2021).
\bibitem{PK2022} P. Kongkhambut, J. Skulte, L. Mathey, J. G. Cosme, A. Hemmerich, and H. Ke{\ss}ler, Observation of a Continuous Time Crystal, Science {\bf 377}, 670 (2022).
\bibitem{TL2023} T. Liu, J. Y. Ou, K. F. MacDonald, and N. I. Zheludev, Photonic metamaterial analogue of a continuous time crystal, Nat. Phys. {\bf 19}, 986 (2023).
\bibitem{VM2023} V. Montenegro, M. G. Genoni, A. Bayat, and M. G. A. Paris, Quantum metrology with boundary time crystals, Commun. Phys. {\bf 6}, 304 (2023).
\bibitem{XW2024} X. Wu, Z. Wang, F. Yang, R. Gao, C. Liang, M. K. Tey, X. Li, T. Pohl, and L. You, Dissipative time crystal in a strongly interacting Rydberg gas, Nat. Phys. {\bf 20}, 1389 (2024).
\bibitem{FR2025_2} F. Russo, and T. Pohl, Quantum Dissipative Continuous Time Crystals, Phys. Rev. Lett. {\bf 135}, 110404 (2025).
\bibitem{ZW2025_2} Z. Wang, R. Gao, X. Wu, B. Bu\v{c}a, K. M{\o}lmer, L. You, and F. Yang, Boundary Time Crystals Induced by Local Dissipation and Long-Range Interactions, Phys. Rev. Lett. {\bf 135}, 230401 (2025).
\bibitem{JW2026} J. Wang, S. Yang, Z. Wang, R. Qi, H. Hu, W. Li, and J. Jie, Non-Resonant Boundary Time Crystals from Quantum Synchronization Breakdown, arXiv: 2603. 14311 (2026).
%%% heat
\bibitem{MR2017} M. Reitter, J. N\"{a}ger, K. Wintersperger, C. Str\"{a}ter, I. Bloch, A. Eckardt, and U. Schneider, Interaction Dependent Heating and Atom Loss in a Periodically Driven Optical Lattice, Phys. Rev. Lett. {\bf 119}, 200402 (2017).
\bibitem{AR2020_2} A. Rubio-Abadal, M. Ippoliti, S. Hollerith, D. Wei, J. Rui, S. L. Sondhi, V. Khemani, C. Gross, and I. Bloch, Floquet Prethermalization in a Bose-Hubbard System, Phys. Rev. X {\bf 10}, 021044 (2020).
%%% decoherence
\bibitem{NQ2016} N. Q. Burdick, Y. Tang, and B. L. Lev, Long-Lived Spin-Orbit-Coupled Degenerate Dipolar Fermi Gas, Phys. Rev. X {\bf 6}, 031022 (2016).
%%% noise
\bibitem{TA1997} T. A. Savard, K. M. O'Hara, and J. E. Thomas, Laser-noise-induced heating in far-off resonance optical traps, Phys. Rev. A {\bf 56}, R1095 (1997).
\bibitem{ML2022} M. L. Day, P. J. Low, B. White, R. Islam, and C. Senko, Limits on atomic qubit control from laser noise, npj Quantum Inf. {\bf 8}, 72 (2022).
%%% drift
\bibitem{JM2019} J. M. Robinson, E. Oelker, W. R. Milner, W. Zhang, T. Legero, D. G. Matei, F. Riehle, U. Sterr, and J. Ye, Crystalline optical cavity at 4K with thermal-noise-limited instability and ultralow drift, Optica {\bf 6}, 240 (2019).
\bibitem{SJ2025} S. Johnson, S. Mishra, A. Pathak, and S. De, Frequency drift corrected ultra-stable laser through phase-coherent fiber producing a quantum channel, Commun. Phys. {\bf 8}, 508 (2025).
%%% SM
\bibitem{SM} See the Supplemental Material for detailed derivation of the mean field approximation, the semi-classical analysis, the complementary
quantum evidence for the BTC and the implementation in a waveguide-QED setup, which includes Refs.~\cite{FI2018x,AC1866x,HW1931x,EP1959x}.

%%% CG
\bibitem{AC1866x} A. Clebsch and P. Gordan, \textit{Theorie der Abelschen Funktionen} (Teubner, Leipzig, 1866).
\bibitem{HW1931x} H. Weyl, \textit{The Theory of Groups and Quantum Mechanics} (Dover, New York, 1931).
\bibitem{EP1959x} E. P. Wigner, \textit{Group Theory and Its Application to the Quantum Mechanics of Atomic Spectra} (Academic Press, New York, 1959).
%%% JC coupling
\bibitem{ET1963} E. T. Jaynes, and F. W. Cummings, Proc. IEEE {\bf 51}, 89 (1963).
%%% limit cycle and quantum diffusion
\bibitem{SD2025} S. Dutta, S. Zhang, and M. Haque, Quantum Origin of Limit Cycles, Fixed Points, and Critical Slowing Down, Phys. Rev. Lett. {\bf 134}, 050407 (2025).
%%% spectral closure insufficient
\bibitem{HA2022} H. Alaeian and B. Bu\v{c}a, Exact multistability and dissipative time crystals in interacting fermionic lattices, Commun. Phys. {\bf 5}, 318 (2022).
%%% two-time corrlation
\bibitem{KT2018} K. Tucker, B. Zhu, R. J. Lewis-Swan, J. Marino, F. Jimenez, J. G. Restrepo, and A. M. Rey, Shattered time: can a dissipative time crystal survive many-body correlations?, New J. Phys. 20, 123003 (2018).
\bibitem{TC2022} T.-C. Guo and L. You, Quantum phases of time order in many-body ground states, Front. Phys. 10, 847409 (2022).
%%% quantum regression theorem
\bibitem{MO1997} M. O. Scully and M. S. Zubairy, \textit{Quantum optics} (Cambridge university press, 1997).
%%% TWA
\bibitem{JH2021} J. Huber, P. Kirton, and P. Rabl, Phase-space methods for simulating the dissipative many-body dynamics of collective spin systems, SciPost Phys. \textbf{10}, 045 (2021).


\end{thebibliography}

\begin{thebibliography}{99}
%%% BTC
\bibitem{FI2018} F. Iemini, A. Russomanno, J. Keeling, M. Schir\`{o}, M. Dalmonte, and R. Fazio, Boundary Time Crystals, Phys. Rev. Lett. {\bf 121}, 035301 (2018).
%%% CG
\bibitem{AC1866} A. Clebsch and P. Gordan, \textit{Theorie der Abelschen Funktionen} (Teubner, Leipzig, 1866).
\bibitem{HW1931} H. Weyl, \textit{The Theory of Groups and Quantum Mechanics} (Dover, New York, 1931).
\bibitem{EP1959} E. P. Wigner, \textit{Group Theory and Its Application to the Quantum Mechanics of Atomic Spectra} (Academic Press, New York, 1959).
\end{thebibliography}
\end{document}